\documentclass[conference]{IEEEtran}

\newcommand{\Comment}[1]{}

\usepackage{amssymb}
\usepackage{amsfonts}
\usepackage{amsmath}
\usepackage{algorithm}
\usepackage{algorithmic}
\usepackage{ulem}
\usepackage{color}
\usepackage{verbatim}
\usepackage{multirow}
\usepackage{graphicx}
\usepackage{graphics}
\usepackage{balance}
\usepackage{mathtools}
\usepackage{pgfplots}
\usepackage{caption}
\usepackage{subcaption}
\usepackage{rotating}
\usepackage{xspace}
\usepackage{algorithm}
\usepackage{algorithmic}

\def\R{\rm I\!R}

\def\mySpace{-0.3cm}

\newcommand{\offset}{\textsc{Offset}\xspace}

\newcommand{\argmin}{\operatornamewithlimits{argmin}}

\Comment{
\newcommand{\orenIeee}[1]{\noindent{\textcolor{blue}{\{{\bf OrenIeee:} {\em #1}\}}}}
\newcommand{\naamaIeee}[1]{\noindent{\textcolor{red}{\{{\bf NaamaIeee:} {\em #1}\}}}}
\newcommand{\rotemIeee}[1]{\noindent{\textcolor{magenta}{\{{\bf RotemIeee:} {\em #1}\}}}}
\newcommand{\yohayIeee}[1]{\noindent{\textcolor{orange}{\{{\bf YohayIeee:} {\em #1}\}}}}
}
\newcommand{\orenIeee}[1]{}
\newcommand{\naamaIeee}[1]{}
\newcommand{\rotemIeee}[1]{}
\newcommand{\yohayIeee}[1]{}

\Comment{
\newcommand{\orenNew}[1]{\noindent{\textcolor{blue}{\{{\bf OrenNew:} {\em #1}\}}}}
\newcommand{\naamaNew}[1]{\noindent{\textcolor{red}{\{{\bf NaamaNew:} {\em #1}\}}}}
}
\newcommand{\orenNew}[1]{}
\newcommand{\naamaNew}[1]{}

\Comment{
\newcommand{\orenFinal}[1]{\noindent{\textcolor{blue}{\{{\bf OrenFinal:} {\em #1}\}}}}
}
\newcommand{\orenFinal}[1]{}

\Comment{
\newcommand{\oren}[1]{\noindent{\textcolor{blue}{\{{\bf Oren:} {\em #1}\}}}}

\newcommand{\naama}[1]{\noindent{\textcolor{red}{\{{\bf Naama:} \em #1\}}}}

}
\newcommand{\oren}[1]{}
\newcommand{\naama}[1]{}

\Comment{
\newcommand{\orenIeeeFinal}[1]{\noindent{\textcolor{blue}{\{{\bf orenIeeeFinal:} {\em #1}\}}}}
\newcommand{\naamaIeeeFinal}[1]{\noindent{\textcolor{red}{\{{\bf naamaIeeeFinal:} {\em #1}\}}}}
}
\newcommand{\orenIeeeFinal}[1]{}
\newcommand{\naamaIeeeFinal}[1]{}

\begin{document}

\title{Audience Prospecting for Dynamic-Product-Ads in\\ Native Advertising}

\author{
  \IEEEauthorblockN{E. Abutbul, Y. Kaplan, N.~Krasne, O.~Somekh}
  \IEEEauthorblockA{
    \textit{Yahoo Research}, Haifa, Israel\\
    \{eabutbul,yohay,naamah,orens\}@yahooinc.com
  }
  \and
  \IEEEauthorblockN{O.~David, O.~Duvdevany, E.~Segal}
  \IEEEauthorblockA{
    \textit{Tech Yahoo}, Ramat-Gan, Israel\\
    \{or.david,omer.duvdevany,evgenys\}@yahooinc.com
   } 
 }

 \IEEEoverridecommandlockouts\IEEEpubid{\makebox[\columnwidth]{979-8-3503-2445-7/23/\$31.00~\copyright~2023 IEEE \hfill} \hspace{\columnsep}\makebox[\columnwidth]{ }}

\maketitle

\begin{abstract} 
With yearly revenue exceeding one billion USD, Yahoo Gemini native advertising marketplace serves more than\orenNew{same as the introduction now} two billion impressions daily to hundreds\rotemIeee{of}\orenIeee{done} of millions of unique users. One of the fastest growing segments of Gemini native is \textit{dynamic-product-ads} (DPA), where major advertisers, such as Amazon and Walmart, provide catalogs with millions of products for the system to choose from and present to users. The subject of
this work is finding and expanding the right audience for each DPA ad, which is one of the many challenges DPA presents. Approaches such as targeting various user groups, e.g., users who already visited the advertisers' websites (\textit{Retargeting}), users that searched for certain products (\textit{Search-Prospecting}), or users that reside in preferred locations (\textit{Location-Prospecting}), have limited audience expansion capabilities.

In this work we present two new approaches for audience expansion that also maintain predefined performance goals. The \textit{Conversion-Prospecting} approach predicts DPA conversion rates based on Gemini native logged data,\rotemIeee{which part is the former?}\orenIeee{done} and calculates the expected \textit{cost-per-action} (CPA) for determining users' eligibility to products and optimizing DPA bids in Gemini native auctions. To support new advertisers and products, the \textit{Trending-Prospecting} approach matches trending products to users by learning their tendency towards products from advertisers' sites logged events. The tendency scores indicate the popularity of the product and the similarity of the user to those who have previously engaged with this product.
\yohayIeee{I don't think that we need details about offset in the above para to be in the abstract.}\orenIeee{done}

The two new prospecting approaches were tested online, serving real Gemini native traffic, demonstrating impressive DPA delivery and DPA revenue lifts while maintaining most traffic within the acceptable CPA range (i.e., performance goal). In particular, we measure $10\%$ delivery lift and $1.58\%$ revenue lift for \textit{Conversion-Prospecting}, and almost $13\%$ delivery lift and $6.33\%$ revenue lift for \textit{Trending-Prospecting}. After a successful testing phase, the proposed approaches are currently in production and serve all Gemini native traffic.\rotemIeee{consider changing to "the proposed approaches are currently in production and serve all Gemini native traffic"}\orenIeee{done}
\end{abstract}

\section{Introduction}\label{sec: introduction}
Yahoo Gemini native marketplace\footnote{See https://gemini.yahoo.com/advertiser/home} serves users with ads that are rendered to resemble the surrounding web\rotemIeee{is the capital W in pupose?}\orenIeee{done} page content. Operating with a yearly run-rate of more than one billion USD\footnote{For commercial confidentiality reasons, we provide approximate estimates and relative figures for traffic volume, revenue, and performance metrics.}, Gemini native is one of Yahoo's\rotemIeee{'s}\orenIeee{done} main and fastest\rotemIeee{fastest?}\orenIeee{done} growing businesses. With more than two billion impressions daily, and an inventory of a several hundred thousand active ads, this marketplace performs real-time \textit{first-price} auctions that accounts for budget considerations and targeting.

\begin{figure}[t]
\centering
\includegraphics[width=0.8\columnwidth]{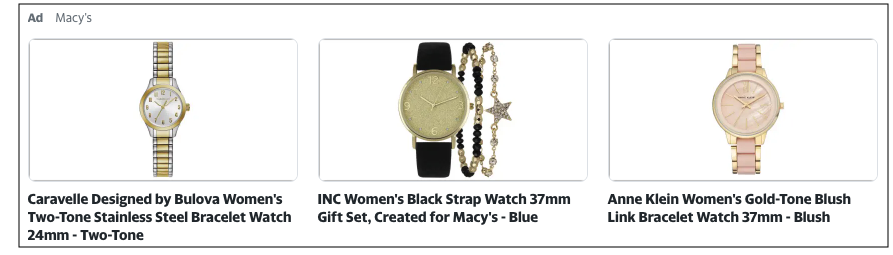}
\caption{A typical Gemini native DPA Carousel Ad.\oren{new: can we find less creepy DPA ad?}\naama{Is this better?}}
\label{fig: carousel}
\end{figure}

\textit{Dynamic-product-ads (DPA)} is a type of online advertising specifically designed for e-commerce merchants, such as Amazon, Walmart and Macy's \cite{bai2022improving}. The merchants provide their product catalogs (usually containing millions of products) and the system automatically creates a personalized ad experience for the users by choosing products from the provided catalogs (see Fig. \ref{fig: carousel}). DPA supports two main types of campaigns, (a) \textit{Retargeting} serves users that already showed an interest in products at the advertiser website. In cases\rotemIeee{such cases?}\orenIeee{fixed} where such users visit websites supported by DPA, those products may be presented to them as a re-targeted DPA ad; (b) \textit{Prospecting} strives to expand the advertisers' audience by presenting products to users that haven't\orenIeee{necessarily} directly interacted with those products\rotemIeee{them}\orenIeee{fixed}. This is achieved by using various DPA types eligibility models that leverage additional signals such as the users' search history or their location.

This work introduces two novel methods for audience expansion (or prospecting) that utilize users' and products' features, and offer\rotemIeee{offers} a less restrictive expansion approach compared to those of other prospecting types (such as Search and Location-Prospecting). The first method\rotemIeee{strives to achieve?}\orenIeee{done} strives to achieve a pre-calculated \textit{cost-per-action} (CPA) goal, by predicting DPA conversion rates and using the latter to calculate the expected CPA for determining users' eligibility and optimizing DPA bids in Gemini native\rotemIeee{Native with capital N}\orenIeee{I write Gemini native} auctions. The predictions are provided by a variant of \offset\footnote{The reader is referred to Appendix \ref{sec: offset} for more details regarding the \offset algorithm.} -- a feature enhanced \textit{collaborative-filtering} (CF) based event prediction algorithm powering many models in Gemini native \cite{aharon2013off}\cite{aharon2017adaptive}\cite{aharon2019soft}\cite{arian2019feature}\cite{kaplan2021dynamic}\cite{kaplan2021unbiased}. The new \textit{Conversion-Prospecting} approach was tested in an online environment, serving real Gemini native traffic, and demonstrated a significant $10\%$ lift in DPA delivery (i.e., more impressions)\rotemIeee{what does lift in audience mean?}\orenIeee{more traffic}\yohayIeee{lift in audience sounds like an increase in the targeted audience size. Do you mean lift in impression delivery?}\orenIeee{fixed} and $1.58\%$ lift in DPA revenue, while maintaining the CPA within the acceptable range (defined w.r.t the CPA of the \textit{Retargeting} DPA type). 
The \textit{Trending-Prospecting} approach supports new advertisers\rotemIeee{consider changing to "a wider range of advertisers"}, and matches trending products to users by learning user-product tendencies from logged events in the advertisers' sites, and random events sampled from Gemini native logged data. The tendency scores are also provided by a variant of \offset, indicating the popularity of a product and the similarity of the user to those previously engaged with the product.
This new approach was also tested in an online environment, serving real Gemini native traffic, and demonstrating almost\rotemIeee{remove the "a"}\orenIeee{done} $13\%$ lift in DPA delivery and $6.33\%$ lift in DPA revenue, while maintaining the CPA within the acceptable range\orenIeee{we should state the acceptable range}\naamaIeee{We can state the 10\% we mention later}.\orenIeeeFinal{change "push to production"}\orenIeeeFinal{done} After a successful testing phase, the proposed approaches are deployed in production, serving all Gemini native traffic. Moreover, the two new approaches now account for over 80\% of the DPA prospecting traffic of advertisers that share their conversion data with Yahoo.\rotemIeee{rephrase this sentence}\orenIeee{done}\yohayIeee{last sentence still isn't clear. do you mean "Moreover, the two new approaches now account for over 80\% of the DPA prospecting impressions/revenue"}\orenIeee{partially done} 
\Comment{
and demonstrated a significant $10\%$ lift in DPA audience and $1.58\%$ lift in DPA revenue, while maintaining the CPA within the acceptable range (defined w.r.t the CPA of the Retargeting audience expansion approach). 
 After a successful testing phase, the proposed approach was pushed to production and currently serves all Gemini native traffic. 
 }

The main contributions of this work are:
\begin{itemize}
    \item We provide a comprehensive overview of the Yahoo Gemini DPA system.
    \item We introduce two new DPA prospecting types, (a) \textit{Conversion-Prospecting}, which discovers an audience that is more likely to converge; and (b) \textit{Trending-Prospecting}, which discovers an audience for new advertisers and their trending products that do not necessarily have purchasing history in Yahoo properties.\rotemIeee{this is the only point that's an actual contribution}
    \item The new DPA prospecting types were successfully tested in web scale, serving real Gemini native traffic, and demonstrating an impressive audience expansion capabilities while maintaining predefined performance goals.\rotemIeee{this and the next point are not contributions}
    \item The new DPA prospecting types are currently\orenIeeeFinal{omit "fully" and save a line}\orenIeeeFinal{done} deployed in production, serving all Yahoo Gemini native traffic.
\end{itemize}

\naamaNew{Since we have space I'm adding how the paper is organized}

The rest of the paper is organized as follows. In Section \ref{sec:Related work} we review related work. In Section \ref{sec: DPA System} we describe the DPA system. Our new\yohayIeee{why is prospecting capitalized? I would just change it but I see it in other places in the paper as well} prospecting approaches are presented in Section \ref{sec:our approach}. In Section \ref{sec:evaluation} we present the performance evaluation results. We conclude and consider future work in Section \ref{sec:Concluding remarks}. \offset\ --\ Yahoo proprietary event prediction algorithm description, is included in the Appendix.
\vspace{-0.1cm}

\section{Related work}\label{sec:Related work}
In general, audience prospecting in online DPA systems has not been significantly covered in previous work\rotemIeee{consider rephrasing to "In general, DPA has not been studied much in the past" }\orenIeee{partially done}\yohayIeee{There has been a lot of work on e-commerce. For instance https://arxiv.org/abs/1606.07154 has 360 citations. do you mean that some specific aspect hans't been studied much?}\orenIeee{I would say tha DPA is a specific type of online advertising or e-commerce}. 
However, in a recent paper \cite{bai2022improving}, the authors present\rotemIeee{presented} a new \textit{similar-product-recommendation} approach used for \textit{Out-of-Stock} scenarios, which serves in Gemini native DPA along with the new prospecting approaches presented in this paper. In particular, the authors use a \textit{Siamese Transformer-based model} to retrieve similar products and then refine them with the attribute \textit{product-name} that indicates the product type (e.g., running shoes, engagement ring, etc.) for post filtering. This is an improvement to the \textit{Deep Siamese model} \cite{reimers2019sentence}, that allows efficient retrieval but does not put enough emphasis on key product attributes.

DPA prospecting is a specific type of audience expansion technology that targets and identifies new audiences with similar attributes of the original (or seed) target audience. Methods to achieve audience expansion are considered in several works, such as \cite{liu2016audience} that describes the campaign expansion approach being used in LinkedIn, and the Hubble system \cite{zhuang2020hubble} used for mobile marketing. 

Next we focus on \textit{conversion-rate} (CVR) prediction, which is a key technology here. One of the early works that discusses the difficulties of predicting \textit{pay-per-action}, such as conversions, is found in \cite{mahdian2007pay}. The authors in \cite{agarwal2010estimating} present a logistic model to determine click and conversion rates. The prediction of post-click conversion rates for display ads is addressed in \cite{rosales2012post} and \cite{chapelle2015simple}. In \cite{yang2016large}, the authors utilize \textit{transfer learning} between campaigns to predict the CVR of display ads. See also \cite{lu2017practical} for a recent study which highlights the
impact of over-prediction and over-bidding involved with CVR prediction, and offers a ``safe'' prediction framework with conversion attribution adjustments to address these issues.

Recommendation technologies play a crucial role in predicting \textit{click-through-rate} (CTR) and CVR, and make it easier for users to find what they want on the Internet. \textit{Collaborative-filtering} (CF) and in particular, \textit{matrix-factorization} (MF) based approaches are regarded as leading recommendation technologies. These approaches represent entities using latent vectors that are learned from user feedback such as ratings, clicks, and purchases \cite{koren2009matrix}. CF-MF based models have been applied successfully in various recommendation tasks such as movie recommendations \cite{bell2007lessons}, music recommendations \cite{aizenberg2012build}, ad matching \cite{aharon2013off}, and more. CF-MF is continuously evolving, with recent advancements combining it with \textit{deep-learning} (DL) for entity embedding \cite{he2017neural}.\orenNew{Maybe add/replace with a more recent reference}

\orenNew{May be omitted} The \textit{Factorization Machines} (FM) model family encompasses many recent CF-MF architectures \cite{rendle2010factorization}.
For instance, \offset\ -- Yahoo's\rotemIeee{'s}\orenIeee{done} proprietary event prediction algorithm (see Appendix \ref{sec: offset}), is closely related to \textit{Field-aware Factorization Machines} (FFM) \cite{juan2017field}\cite{juan2016field} and \textit{Field Weighted Factorization Machines} (FwFM) \cite{pan2018field}\naama{how can we fix the line?}\oren{done}. 
However, \offset differs from FFM and FwFM in that the ad side field is unique and entry-wise multiplied by the user features overlapping and independent latent factor vectors, resulting in a sum of triplet and pair multiplications, as opposed to just pair multiplications in FFM and FwFM.\rotemIeee{I suggest changing the order of the related works, start with the last two paragraph, continue with CVR, and end with DPA}

\section{Yahoo Gemini native DPA System}\label{sec: DPA System}

In this section we provide a comprehensive description of our DPA system. We characterize the major DPA types, describe the DPA click model that drives the system, and consider the DPA serving (or inference) system that periodically gets all relevant models and rank the DPA products in real-time with low latency before pushing the top product to Gemini native auctions.

\subsection{Overview}
\begin{figure}[t]
\centering
\includegraphics[width=0.9\columnwidth]{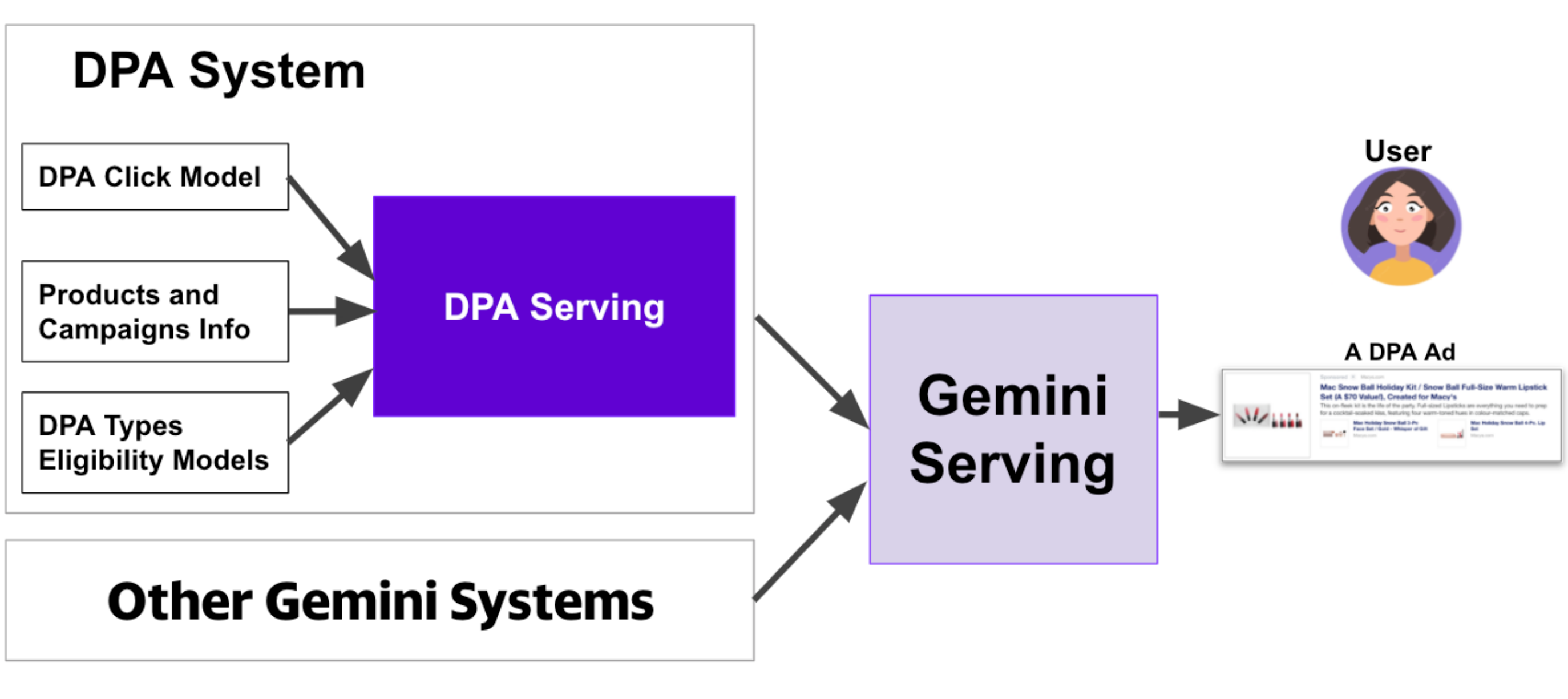}
\vspace{-0.2cm}
\caption{Yahoo Gemini native DPA system components: (a) DPA product click prediction model; (b) DPA Product/Campaign database; (c) DPA types eligibility models; and (d) DPA Serving.}\orenIeee{add one more ranker to the figure so no one thinks DPA is the only ranker}\naamaIeee{done}
\label{fig: DPA components}
\end{figure}

\rotemIeee{consider changing the section name to "The Yahoo Gemini DPA System}
\rotemIeee{you need a connecting sentence to start this section, instead of directly delving into the technical details}\orenIeee{done} The four main components of the Yahoo Gemini DPA system are depicted in Fig. \ref{fig: DPA components}\orenNew{in the figure "dpa types" or "eligibility models", not both (?)}\orenNew{it's correct}: (a) \textbf{DPA click prediction model} predicts the probability for a certain user to click on a certain product (see Section \ref{subsec:DPA pctr}); (b) \textbf{DPA database} contains\rotemIeee{containing} products and campaigns' information which is\rotemIeee{are} required for filtering products (e.g., campaign budget, campaign targeting, etc.)\rotemIeee{I suggest to remove all the examples in the brackets, there's too many brackets with the section references as it is, and you will elaborate on this anyway later}, getting the advertisers' bids, and fetching the products' assets (e.g., image, title, and description) for rendering the ads; (c) \textbf{DPA types eligibility models}\naamaIeeeFinal{Consider adding "capture"}\orenIeeeFinal{done (sort of)} includes the various DPA types (see Section \ref{subsec:DPA types})\rotemIeee{remove "various DPA types" to match the other component descriptions}, where each type has its unique logic for determining which products are eligible\rotemIeee{remove "eligible", it makes no sense with it} good candidates for presentation to the incoming users; and (d) \textbf{DPA serving system} uses the above three components for selecting and rendering the top DPA ads to compete in Gemini native \textit{first-price} auctions (see Section \ref{subsec:DPA serving}). 

\vspace{-0.1cm}
\subsection{DPA types}\label{subsec:DPA types}
In Gemini native DPA there are\rotemIeee{consider using "there are" instead of "we have"}\orenIeee{done} two types of campaigns the advertiser can choose from,\yohayIeee{no need for the "i.e." here} Retargeting and Prospecting.\orenFinal{it's different than the description at the Introduction} \textbf{Retargeting campaigns} may present a product to a user only if there was a direct interaction between the user and a sufficiently similar product on the advertiser's site\yohayIeee{wouldn't it be more accurate to sat that retargeting requires the user to have had a previous interaction with the specific product or products like it?}. \textbf{Prospecting campaigns} allow various audience expansions approaches (or DPA types eligibility models)\naamaNew{or eligibility DPA models} to\rotemIeee{maybe "and" instead of "to"?} determine \yohayIeee{verify implies a "true" state that we're checking, maybe evaluate/determine instead?} the eligibility of incoming users to the campaigns. Due to commercial confidentiality matters, only a partial list of the supported DPA types are described below.

\paragraph{Retargeting types}
\begin{itemize}
    \item {\textbf{Retargeting}} suggests products based on the users' past engagements (e.g., \textit{view}, \textit{add-to-cart}, and \textit{purchase}) at the advertisers' sites.\oren{is that correct?}\naama{yes-looks good, it can also be purchase but it depends on the product}
    \Comment{targets the users' products and recommendations based on their purchasing history.\oren{is that correct?}}
    \item {\textbf{Cross-Sell}} suggests complementary products based on the users' purchasing history.
    \item \textbf{Out-of-Stock} recommends alternative products for \textit{out-of-stock} re-targeted products \cite{bai2022improving}. \naama{new, please revisit}\oren{done}
\end{itemize}
\paragraph{Prospecting types}
\begin{itemize}
    \item {\textbf{Search-Prospecting}} recommends relevant products based on the users' search phrases history.
    \item \textbf{Location-Prospecting} products are matched to users based on their\rotemIeee{logged}\orenIeee{done} logged \textit{Where On Earth Identifier} (WOEID) locations. This is done through a feed provided by the advertiser, which specifies the connection between products and the corresponding WOEIDs. 
\end{itemize}
As mentioned earlier, in this work we present two new DPA prospecting types that are now available to our advertisers.
\begin{itemize}
    \item \textbf{Conversion-Prospecting} the system predicts the probabilities that users will convert on products (i.e, purchasing) and uses these predictions\rotemIeee{there is no former and latter in this sentence...}\orenIeee{done} for determining the eligible products for the users, and for optimizing the DPA bids in Gemini native auctions. For more details see Section \ref{sec:convPros}.
    \item {\textbf{Trending-Prospecting}} recommends new trending products based on the users' features by learning their preferences from logged events in the advertisers' sites. For more details see Section \ref{sec: trendy}.
\end{itemize}

\Comment{It is noted that DPA includes also \textit{Broad-Audience} campaigns. These campaigns support both Retargeting and Prospecting goals, and the system automatically controls which DPA type is eligible for each user and what is the optimal bid to be used. It is considered as the ``next step'' of DPA, where advertisers are oblivious to the various DPA types and is outside the scope of this work.\naama{about 35\% of the spend is work on this manner, it should be in future tense.}}
\vspace{-0.2cm}
\subsection{DPA click prediction model}\label{subsec:DPA pctr}
This key component in the DPA system is used for generating the predicted click probability $\mathrm{pCTR_{u,p}}$ of a user $u$ to click on product $p$\rotemIeee{consider changing to "user u on product p}\orenIeee{partially done}. To train the model we use a variant of the \offset event prediction algorithm (see Appendix \ref{sec: offset}). Here, unlike most models driven by \offset, we consider products instead of native ads\rotemIeee{consider rephrasing to "we consider products instead of native ads"}\orenIeee{done} (see Fig. \ref{fig: carousel} for\rotemIeee{an example of}\orenIeee{done} an example of a DPA carousel ad with three products). Following are short descriptions of the user features, product features, and user-product similarity features, currently used by the DPA click prediction model\rotemIeee{consider rephrasing whole sentence to "Our production model currently supports the following features:"}.
\paragraph{User features}
\begin{itemize}
    \item \textit{techno-segments}\rotemIeee{consider adding ":" to the end of each feature} a multi-value feature (see \cite[Section 3.1.1]{arian2019feature})
    that provides technical details such as the user's device type, operating system, browser, Internet provider, and carrier.
    \item \textit{page-section} a categorical feature indicating the web page id where the user viewed the product ad. \naama{copied from conv-pros features}\oren{looks good}
    \Comment{the Web page id, with several thousands values.}
    \item \textit{dpa-type-experiment-id} a categorical feature\rotemIeee{for some features you specify that it's categorical, but not for others. Is this specification really important?} denoting the DPA type and experiment used for serving the user. As described in Section \ref{subsec:DPA types} we have many different DPA types, each of them can be served by\rotemIeee{"can be served by"}\orenIeee{done} different experiments that test new capabilities and\rotemIeee{that test new capabilities. Remove the "that performs differently" part}\orenIeee{partially done} perform differently. \naama{I extended the explanation to experiment id}\oren{looks good}
    \item \textit{impression-history} a categorical feature representing how many times the user has seen\rotemIeee{saw} Gemini native ads during the past 30 days, divided into several bins.\orenIeeeFinal{replace "that represents" with "representing"}\orenIeeeFinal{replace "last month" with "past 30 days"}\orenIeeeFinal{replace "in" with "during"}\orenIeeeFinal{all done}\naama{I think I wrote the wrong explanation, it shouldn't be product but the num of native ads}\oren{looks good}
    \item \textit{age} the user's age divided into 5 year bins.
    \item \textit{user-clicked-category} a multi-value feature denoting the ad categories that the user has clicked on during the past two weeks.
    \item \textit{mobile-activity} a multi-value feature\orenIeeeFinal{replace "that represents" with "representing"}\orenIeeeFinal{done} representing the user's mobile app activity history.
    \item \textit{ctr-advertiser-top} a multi-value feature holding a list of advertisers with the user's highest CTR, during the last two week.\oren{do you know what size is the list?}
    \Comment{for each user we keep a list of the advertisers that the user has the highest CTR during the last two week.
    \naama{I've changed the explanation similarly to what you wrote on ctr-campaign-top}
    \Comment{for each user we keep a list of the advertisers that the user has the highest probability to click on in the last two weeks.}\oren{new: this explanation is bad. Let's discuss it}\oren{got it}}
    \item \textit{user-clicked-product-category} a multi-value feature holding a list of the user's clicked products' categories during the past week.
\end{itemize}
\paragraph{Product features} We use a hierarchy of self-explanatory ad features:\orenIeee{I removed "default" to avoid confusion with "default-product-group"} 
\textit{advertiser-id} $\rightarrow$ \textit{product-set-id} $\rightarrow$ \textit{product-group-id} $\rightarrow$ \textit{product-id} and add all relevant vectors as mentioned in Appendix \ref{sec: offset} to produce the final product latent factor (LF) vector. Since there are millions of products and we want to keep a reasonable model size\rotemIeee{due to latency constraints}, we train only\rotemIeee{consider replacing with "we train only..."}\orenIeee{done} on products that have more than a predefined number of impressions so far (e.g., 1000). For all other products we train a \textit{default-product-group} vector that is included under each \textit{product-group-id} value\rotemIeee{what does that mean?}. We also use the \textit{default-product-group} vector\rotemIeee{consider using "default vector" instead of latter, it's confusing in this context}\orenIeee{done} to initialize the \textit{product-id} vector once it has sufficient\rotemIeee{number of}\orenIeee{done} number of impressions.

\paragraph{Similarity features} These user-product features are used as\rotemIeee{"used as" or "translated to"}\orenIeee{done} weights that are added to the\rotemIeee{the}\orenIeee{done} \offset score (see Eq. \eqref{eq: score}\rotemIeee{of the appendix}) and trained as \textit{logistic regression} parameters \cite{aharon2019soft}.
\begin{itemize}
\item \textit{frequency} represents the number of times the user has seen an ad product from the same campaign during the past week, divided into several bins\rotemIeee{is it necessary to mention that it's binned?}.\orenNew{seen the product at the advertiser's site or seen the product ad at Yahoo?}\naamaNew{Seen the product ad at Yahoo}\orenNew{done}
\item \textit{recency} represents the time period since the user has last seen the product at the advertiser's site divided into several bins.\naama{Write as a binning feature}\oren{it's ok to cheat here}
\item \textit{slot-device} represents the slot index (relevant to DPA carousel ads\rotemIeee{there is no explanation on what carousel ads are. Is there a paper to reference? Maybe add an example to figure 1}\orenIeee{done}. See Fig. \ref{fig: carousel} for a DPA carousel ad with 3 slots) and whether the device is mobile (e.g., ``slot3\_mobile'' or ``slot1\_nonMobile'').
\end{itemize}
\vspace{-0.2cm}
\subsection{DPA serving}\label{subsec:DPA serving}
\begin{figure}[t]
\centering
\includegraphics[width=0.8\columnwidth]{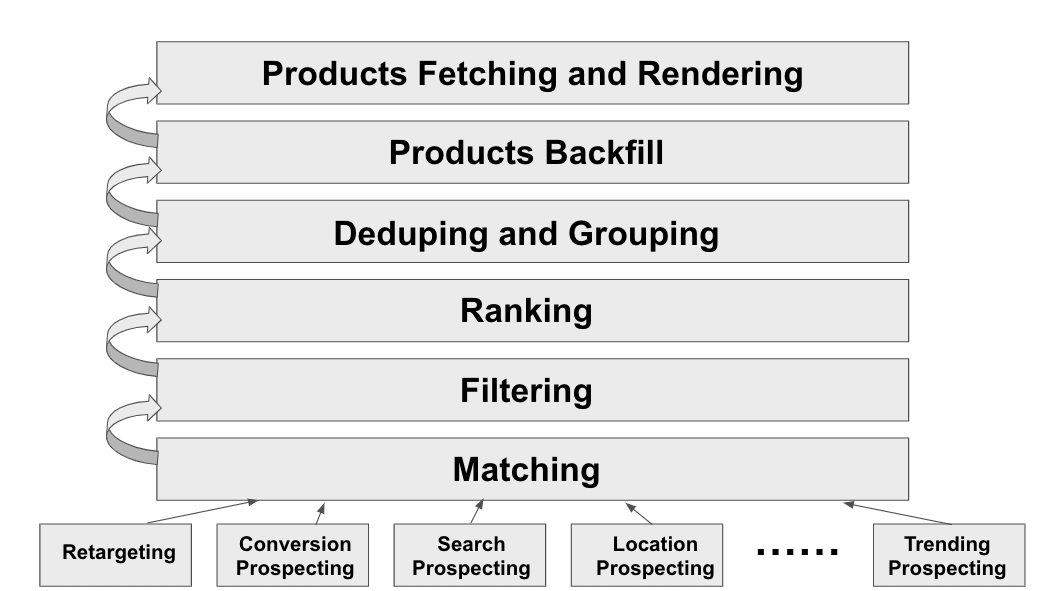}
\vspace{-0.1cm}
\caption{DPA Serving flow.\rotemIeee{the direction of the arrows is confusing, it's reversed from the order of actions you're describing}}
\label{fig: serving flow}
\end{figure}

\yohayIeee{I suggest moving this subsection to before the click prediction one, and maybe before the dpa types one. It would the fist introduce the reader to the broad view of the system before giving more details on the 2 specific elements (click prediction and types)}
The DPA Serving flow, which occurs for every incoming user impression\rotemIeee{consider rephrasing to "The DPA Serving flow for every incoming user..."}, is depicted from bottom to top in Fig. \ref{fig: serving flow}.

\noindent\textbf{Gathering products from the different DPA types}\rotemIeee{consider adding ":" to the end of each item name} each type returns a list of products that are eligible to the user, based on their different algorithms. For example, the \textit{Retargeting} DPA type\rotemIeee{remove ", that"}\orenIeee{done} returns the products from the user's browsing history at the advertisers' sites.

\noindent\textbf{Matching} for each product, this stage returns information on the product ad group and campaign (e.g., budget, target audience, etc.).
\Comment{and builds the product hierarchy: \textit{advertiser} $\rightarrow$ \textit{campaign-id} $\rightarrow$ \textit{product-set-id} $\rightarrow$ \textit{product-group-id}  $\rightarrow$ \textit{product-id}.\orenNew{not sure why we build the hierarchy and why it's different from the hierarchy we mention above. Maybe we can omit the list and just mention it builds the hierarchy}}

\noindent\textbf{Filtering} filters the eligible products that cannot be presented due to various reasons. For example, the user is not included in the advertiser's target audience (e.g., the user gender does not match the campaign targeting), the campaign's expiration date is due, or that there are issues with the publisher's policy (e.g., language).

\noindent\textbf{Ranking} uses the DPA click prediction model (see Section \ref{subsec:DPA pctr}) to get the user-products predicted click probabilities, 
multiply those by the advertisers' bids to score\footnote{Roughly speaking, the DPA score (i.e., the click probability times the advertiser's bid) is the expected revenue from the event of presenting the product to the user.} the products, and rank them accordingly.

\noindent\textbf{Deduping and Grouping} the DPA flow returns one ad per advertiser, by picking the advertiser's ad with the highest score. To do so, two stages are needed. \textbf{Deduping:}\rotemIeee{":"}\orenIeee{done} when the same product is returned from two different DPA types, the type with the lower score (see \textbf{Ranking} stage)\rotemIeee{is this the pCTR score? It's hard to follow even with the footnote}\orenIeee{done} is removed\footnote{The product's scores for various DPA types vary, as the DPA type is among the user features predicting $\mathrm{pCTR_{u,p}}$ (see Section \ref{subsec:DPA pctr}).}.
\textbf{Grouping:}\rotemIeee{":"}\orenIeee{done} in cases the impression web page (or publisher's page)\rotemIeee{just say "when the publisher's page"} supports carousels ads
(See Fig. \ref{fig: carousel})\rotemIeee{"when" something applies "then" something happens. This sentence is missing the "then" part, which is in the next sentence. Please combine them for correct syntax}\orenIeee{done}, the system attempts to group different products from the same \textit{product-group-id} into one carousel ad.\orenIeeeFinal{there is a redundancy here}\orenIeeeFinal{done}

\noindent\textbf{Product backfill} building a carousel ad requires at least three different products. If there is a carousel ad with only one or two products, a call is being made to the \textit{product-recommendation-model} (see Section \ref{sec:Product-recommendation-model}) for filling the vacant carousel slots with adequate products matching the ad pivot product (i.e., the product with the highest score).\oren{is that correct?}\rotemIeee{this explanation is very hard to understand}

\noindent\textbf{Product fetching and rendering} fetching all the relevant products' assets information (e.g., image, title and description) and render the final DPA ad, before sending it to Gemini Serving to compete with other non-DPA ads in the final auction.\rotemIeee{this subsection is called DPA serving, but saying "sending it to Gemini Serving" indicating that it's not really that. Rephrase the sentence to avoid confusion}

\naama{Add an intro on this model / remove it. Or remove the popularity part.}
\subsubsection{Product-recommendation-model}\label{sec:Product-recommendation-model}\naama{new: A new paragraph, please revisit}\rotemIeee{this subsection seems out of place and unrelated}\yohayIeee{agree with Rotem, we can just a very broad single line description to the "product backfill" paragraph}
This model is composed of two sub-models. A CF \textit{Item-to-Item} \cite{linden2003amazon} recommendation model and a product \textit{Popularity} model. If the pivot product doesn't have at least two related products in the Item-to-Item model, the system uses the Popularity model for backfilling the DPA carousel ad; otherwise, it uses the Item-to-Item model.\orenIeeeFinal{it seems the last sentence is redundant}\orenIeeeFinal{changed it according to chatGpt}
\vspace{-0.15cm}

\section{New DPA prospecting types}\label{sec:our approach}
In this section we introduce two new DPA prospecting types (i.e., audience expansion), \textit{Conversion-Prospecting}, and \textit{Trending-Prospecting}. While the latter is aimed for serving new advertisers, the former is designed to maintain a predefined performance goal, i.e., target \textit{cost-per-action} (tCPA).\rotemIeee{is this not just for conversion prospecting?}\orenIeee{done}
\subsection{Conversion-Prospecting} \label{sec:convPros}
\subsubsection{Overview}
\orenNew{much better now. However, we may shorten it}\naamaNew{done}
\Comment{As a result of DPA traffic growth during the last few years, Gemini native gets a considerable amount of DPA conversions (i.e., product purchasing) for many advertisers. Since all product events (DPA and non-DPA) are logged in the system, including impressions, clicks and conversions,}

In this new DPA type, a variant of an \offset-based (see Appendix \ref{sec: offset}) Conversion-Prospecting model is trained to predict the \textit{conversion-given-click} probability $\mathrm{pCONV}_{u,p}$ of user $u$ and product $p$ (see Section \ref{sec: conv-pros model}). During serving time, a final bid (i.e., $\mathrm{bid_{final}}$) is calculated for all products included in the Conversion-Prospecting model. As we shall explain later, the final bid is based on the information provided by the advertiser and the model.\rotemIeee{this reads as "when calculating the bid, the bid is calculated" please revise this sentence}\orenIeee{done} To determine $\mathrm{bid_{final}}$ value, we assume that the following values are given: 
\Comment{During serving time, when setting the bid that product $p$ will use\rotemIeee{"use" instead of "compete"}\orenIeee{done} within the final auction for user $u$, a final bid (i.e., $\mathrm{bid_{final}}$) is calculated \rotemIeee{this reads as "when calculating the bid, the bid is calculated" please revise this sentence} based on information provided by the model and the advertiser.
In order to determine $\mathrm{bid_{final}}$ value, we assume that the following values are given:}
\begin{itemize}
        \item $\mathrm{bid_{product-group}}$ - the advertiser's original bid for products from $p$'s product-group\footnote{Each product belongs to a product-group as part of the \textit{product features} hierarchy, presented in Section \ref{subsec:DPA pctr}.}\rotemIeee{it should be mentioned earlier that a product is part of an ad group. Also, "ad group" is too specific to Gemini, and you've mentioned before that a product is not an ad, so that's extra confusing. Consider a different term}\orenIeee{done}. $\mathrm{bid_{final}}$ should be bounded by this value.\rotemIeee{this sentence doesn't start with a capital letter, consider rephrasing}\orenIeee{done}.
        \item $\mathrm{tCPA_{advertiser}}$ - the target \textit{cost-per-action} of the advertiser. As most advertisers don't provide this value, it is usually \rotemIeee{remove "being"}\orenIeee{done} calculated by the system, as describe in Section \ref{sec:tCPA}.
        \item $\mathrm{pCONV}_{u,p}$ - the predicted probability that user $u$ would convert on a product $p$ after clicking on its ad, provided by the Conversion-Prospecting model (see Section \ref{sec: conv-pros model}).
\end{itemize}
The theoretical expected spend on a click is given by the predicted probability that the user would convert on a product ($\mathrm{pCONV}_{u,p}$), multiplied by how much the advertiser is willing to pay on average\footnote{Recall that according to the \textit{cost-per-click} (CPC) price-type, advertisers are paying for clicks and not for conversions.} for such a conversion ($\mathrm{tCPA_{advertiser}}$)\rotemIeee{this should be a formula on its own}\rotemIeee{which value? what target?}\orenIeee{fixed}.
\Comment{
\orenNew{we should state that each advertiser provides a bid for product-group}\naamaNew{done} A user is eligible for a product, if the potential expected revenue \rotemIeee{which value?} meets the advertiser's target \rotemIeee{what target?}. The expected value \rotemIeee{this should be a formula on its own} is the probability a user would convert on a product ($\mathrm{pCONV}_{u,p}$), multiplied by how much the advertiser is willing to pay for such conversion ($\mathrm{tCPA_{advertiser}}$).}
Moreover, the final bid cannot exceed what the advertiser is willing to pay for a click ($\mathrm{bid_{product-group}}$).
Accordingly, the final bid is set to
\begin{equation}\label{eq:conv pros equi}
\mathrm{bid_{final}} = \min \{\mathrm{pCONV}_{u,p} \cdot \mathrm{tCPA_{advertiser}}\ ,\ \mathrm{bid_{product-group}}\}\ .
\end{equation}
It is apparent, that $\mathrm{bid_{final}}$ has low values for products the user is unlikely to convert on (i.e., purchase), as a result, these products are unlikely to win the auction and be presented to the user. However, in principle a user is eligible to all products for which the $\mathrm{bid_{final}}$ value is above the web page floor price\footnote{A minimum price, typically denominated in US Cents, set by some publishers to restrict auctions.}\rotemIeee{this contradicts the first sentence of the previous paragraph. Also, I don't think there's a need to mention floor price at all}\orenIeee{partially done}. We would like to emphasise, that in contrast to other bidding-strategies, such as \textit{optimized cost-per-click} (oCPC)\rotemIeee{if you don't have a citation from oCPC, mentioning it here is confusing. You're assuming the reader knows what it is and how it works} that sets the bid based on a performance goal (e.g., advertisers' tCPA), according to Conversion-Prospecting, the advertiser provides a bid which is being used as an upper bound to the final bid of Eq. \eqref{eq:conv pros equi}\rotemIeee{not sure how necessary this emphasis is}\orenIeee{added a motivation}. Therefore, in practice $\mathrm{bid_{final}}$ may be much lower and is proportional to the likelihood of the user to convert.

\subsubsection{Calculated target CPA} \label{sec:tCPA}
To determine the user's eligibility to a certain product, we need to obtain the advertiser's performance goal (i.e., $\mathrm{tCPA_{advertiser}}$), which most advertisers are reluctant to share. Nevertheless, our advertisers are willing to pay 50\% more for new users (i.e., prospecting users) than for re-targeted users. Hence, the $\mathrm{tCPA_{advertiser}}$ is set to ($1.5 \cdot \mathrm{CPA_{retarg.}}$), which is calculated for each advertiser beforehand. If the advertiser lacks re-targeting campaigns, we use its prospecting type with the best (i.e., lowest) measured CPA\naamaIeee{We use its prospecting type and not campaign. consider replace lowest with 'best measured CPA'}\orenIeee{done}\orenIeeeFinal{replace "to calculate" with "of calculating"}\orenIeeeFinal{done} for calculating $\mathrm{tCPA_{advertiser}}$. Since Conversion-Prospecting requires conversions to be operational, the whole discussion is valid only for campaigns that report such conversions.
\Comment{However, as most advertisers do not share this information,
the Product team has noted that advertisers are willing to pay 50\% more for new users (i.e., prospecting users) than for re-targeted users, hence the tCPA is set to $(1.5 \cdot CPA_{retarg.})$.  Due to that, we calculate an estimated target CPA for each advertiser beforehand.}

\vspace{-0.05cm}
\subsubsection{Conversion-Prospecting model}\label{sec: conv-pros model}
The conversion model is trained using the \offset event prediction algorithm described in Appendix \ref{sec: offset}. The \offset algorithm\rotemIeee{remove "algorithm" or add "The Offset algorithm)}\orenIeee{done} gets as input mini-batches of new positive and negative events from Gemini native feeds, and incrementally trains its model \rotemIeee{consider rephrasing to "Offset incrementally trains on mini-batches of the latest real positive and negative events from Gemini native"}. 
In our case, the inputs are products' post-click conversions (i.e., products' purchases after the user clicked on a DPA ad) as positive events, and\orenNew{this actually not true. All clicks are negative events}\orenNew{say that all clicks are negative and explain later} clicks as negative events (see more details ahead\rotemIeee{sequel is the wrong word}\orenIeee{done}). 

Due to data sparsity of the events in the product level, the model is not learning the full hierarchy of the product. Instead, the Conversion-Prospecting model includes the following product features only: (1) \textit{product-id}; (2) \textit{product-set-id};
and (3) \textit{advertiser-id}. On the user side, the model features are:\naama{fixed the features list}\oren{fine} (1) \textit{ctr-campaign-top} a multi-value feature holding a list of campaigns with the user's highest CTR, during the last week (similar to \textit{ctr-advertiser-top} of Section \ref{subsec:DPA pctr} but at the campaign level);
(2) \textit{dpa-type-experiment-id} as described in Section \ref{subsec:DPA pctr} we have many different DPA types and experiments, where each of them performs differently; and (3) \textit{page-section} is the Web page id where the user viewed the product ad. 
Since users are represented \rotemIeee{represented}\orenIeee{done} by their features (see Appendix \ref{sec: offset}), audience expansion is less restricted than other prospecting types (such as Search-Prospecting) that require a user id or a bCookie. \orenNew{consider adding the multiplication factor of bid landscaping alpha at the product-group level}

Given a user $u$ and a product $p$, the Conversion-Prospecting model predicts the probability that the user would convert on that product (i.e., purchase the product), $\mathrm{pCONV}_{u,p}$. As with other conversion models in Gemini native, click and conversion events are not joined together due to long conversion delays\footnote{\oren{we should write "post-click conversions"} Conversions may arrive up to 30 days after the user had clicked the ad.}. Therefore, the model under-predicts since it trains on additional negative event for each positive one, and the true prediction is actually $\mathrm{pCONV}_{u,p}\approx\tilde{P}/(1-\tilde{P})$, where $\tilde{P}$ is the ``raw'' prediction.\rotemIeee{it's unclear which P is actually used. Please add this information}\orenIeee{done}

\paragraph{Published model} \rotemIeee{it's weird that a random paragraph has a title when none of the others in this subsection had one} Periodically (e.g., every 6 hours)\naamaIeee{This model runs every 6 hours}, the model is published to the Conversion-Prospecting serving (see Section \ref{sec: Conversion-Prospecting serving})\rotemIeee{which part of the flow? Maybe add it to the drawing?}\orenIeee{done}.
Due to serving limitations, the published model is bounded to $K$ (e.g., $K=1000$) products that are \rotemIeee{remove "being"}\orenIeee{done} picked by taking the top-$K$ products with the highest amount of conversions and that exceed a predefined minimal number of conversions (e.g., 10 conversions). This guarantees that the published model includes products that were trained on a significant amount of events. Moreover, the target \textit{cost-per-action} (i.e., $\text{tCPA}_{\text{advertiser}}$) of all selected products of each advertiser, which is used to determine the final bid (see Eq. \eqref{eq:conv pros equi}),
\rotemIeee{it would be nice to have a diagram of the sequence of events from when a user enters the website -> serve request -> auction sequence -> display to user} is calculated\naamaIeee{"is calculated" or 'used to calculate' should be removed} according to Section \ref{sec:tCPA}, and included in the published model.
\rotemIeee{you said in section DPA Types that the advertiser can choose a prospecting type, and that retargeting is one of them. What happens if they don't have retargeting at all?}\orenIeee{done in section \ref{sec:tCPA}}
\oren{new: keeps it for each product?} \naama{In the published model we add for each product its tCPA, but it's the same val for all products from the same advertiser.}\oren{maybe we should mention earlier that tCPA is calculated for all product of the same advertiser}

\subsubsection{Conversion-Prospecting serving}\label{sec: Conversion-Prospecting serving}
When a new model is published, the system indexes all products' and user features' vectors, and fetches the $\mathrm{bid_{product-group}}$ and $\mathrm{tCPA_{advertiser}}$. During serving time, the serving system uses the published model to predict the \textit{conversion-given-click probability} $\mathrm{pCONV}_{u,p}$ for every product $p$ that is included in the published model and calculate its final bid according to Eq. \eqref{eq:conv pros equi}.
To decrease the load on the serving system, Conversion-Prospecting products are competing in an internal preliminary auction with other prospecting products\naama{add it to trendy publishing as well}. The internal auction is based on a score given for each product, which is the final bid times the predicted click probability (see Section \ref{subsec:DPA pctr}), i.e., $\text{pCTR}_{u,p}\cdot \text{bid}_{\text{final}}$\rotemIeee{should be a formula}\orenIeee{done}. Finally, only the top $L$ (e.g., $L=40$) products continue to the next stage (see Section \ref{subsec:DPA serving}).

\subsection{Trending-Prospecting} \label{sec: trendy}
\yohayIeee{consider re-organizing in a subsequent editing opportunity. information about thepixel feed should be near the beginning}
\subsubsection{Overview}
To serve new advertisers (and their products) that have no event history logged in Gemini native feeds, we introduce the DPA Trending-Prospecting type\rotemIeee{consider rephrasing to "The DPA Trending-Prospecting type serves products of new advertisers without an event history logged in Gemini native"}. This prospecting approach applies a user \textit{lookalike} model (see Section \ref{sec: lookalike model}), which is based on a variant of the \offset algorithm (see Appendix \ref{sec: offset}), to predict the eligibility score (or affinity score)\rotemIeee{is eligibility level really the right description? What about affinity score?}\orenIeee{partially done} of a user to a certain product. Given a user $u$ and a product $p$, the eligibility score $S_{u,p}$ is designed to capture two factors, (a) the popularity of product $p$; and (b) how similar (or lookalike) $u$ is to users that have already engaged\rotemIeee{that have already engaged}\orenIeee{done} with product $p$. A user $u$ would be eligible for\orenIeee{trending product? is this because the Lookalike model is operating only on advertiser logs?}\naamaIeee{how is this relate? We call it trending product because the model pick the top products in the last day} product $p$ if the eligibility score $S_{u,p}$ is above a predefined threshold (see Section \ref{sec: lookalike model}). 
\Comment{I addition, it also affects the training of other users that have already engaged with the product. Moreover, every user-product interaction propagates via the learning process and effects all relevant model parameters.}
\rotemIeee{CPA hasn't been mentioned for this type yet, where does it come from and how is it used?}\orenIeee{done}\rotemIeee{but this is a new advertiser?}\orenIeee{done}

\Comment{
\orenIeee{the following paragraph should be probably omitted}
In order to broaden the audience for new advertisers, we permit a CPA increase of up to $10\%$\rotemIeee{CPA hasn't been mentioned for this type yet, where does it come from and how is it used?}, in comparison to the CPA performance of the advertisers' traffic\rotemIeee{but this is a new advertiser?}\orenIeee{good question. I guess it's for performance evaluation only} over non Trending-Prospecting (see Section \ref{subsec: dpa performance trendy}).
\naamaIeee{Maybe it's best to write: in comparison to the CPA performance of the advertiser where trending-prospecting is not eligible}\orenIeee{we actually don't explain how does Trendy works}
\naamaIeee{done}\orenIeee{I have changed it a bit. please have a look}ֿ\naamaIeee{Thanks looks much better}}

\subsubsection{Lookalike Model}\label{sec: lookalike model}
The Lookalike model uses the \offset algorithm almost without any special modifications. In particular, we use (1) \textit{age}; and (2) \textit{gender} as user features, and (1) \textit{advertiser-id}; (2) \textit{product-set-id}; and (3) \textit{product-id}, as product features (see Section \ref{subsec:DPA pctr}). All events that involve users with unknown age or gender values are ignored\rotemIeee{during the learning phase}. By using the \offset algorithm, the Lookalike model benefits from the advantage of \textit{collaborative-filtering} (CF), i.e., filling information gaps due to sparsity issues by revealing hidden patterns within the data. In particular, a user $u$ that has already engaged with product $p$, contributes to the learning process of all users and products that are sharing similar features.

While the ``classic'' click prediction based \offset algorithm consumes ad clicks and skips (i.e., impressions without clicks)\rotemIeee{best to just write impressions instead of skips and the brackets} data to predict CTR, we alter the algorithm inputs to produce user-product eligibility scores.
\Comment{In principal, the \offset algorithm expects two types of inputs: positive events and negative events. In the ``classic'' click-prediction scenario these are clicks (positives) and skips (negatives).}
Since \offset is designed to produce predictions that approximate the ratio between the number of positive events ({\it \#pos}) to the number of positive and negative events ({\it \#pos+\#neg}), the result is the predicted CTR \rotemIeee{this whole paragraph and the formula should be in the appendix}
\begin{equation}\label{eq: pCTR-Eq}
    \mathrm{pCTR} \approx \frac{\#pos}{\#pos+\#neg} = \frac{clicks}{clicks+skips}\ .
\end{equation}
In the context of the Lookalike model, we train \offset with different inputs. As positive events we take {\bf purchase} and {\bf add-to-chart} events from the external advertisers' pixel feeds\footnote{These are not Gemini native logged data feeds, but feeds provided by the advertisers.}. For efficiency reasons we focus on the top $N$ (e.g., $N=10K$) popular products of the pixel feeds and train only on them. Since the advertisers' pixel feeds do not include negative events,\naamaIeee{Think of adding - since all events on the advertiser's site demonstrate some level of engagement with the products}\orenIeee{are you worried about the "view" event which is no positive?} we use a random sample of size $M$ (e.g., $M=2000$) for each product from the Gemini native impression feed as the negative events for training\rotemIeee{this part is very unclear. Is this taken from the click feed? Does that mean you are taking a fixed number of negative events for each product? Why?}\orenIeee{partially done}\naamaIeee{Maybe we can add - each impression event of this nature simulates a non-interaction event, effectively indicating that the user did not engage with the product}. By doing so we expect \offset to be able to differentiate between users that engaged with a certain product in the advertisers' pixel feeds, and users in the general population, picked by the negative random sample taken from the Gemini native impression feed\rotemIeee{you should explain why click and conv events can't be coupled}.
\Comment{
Figure \ref{fig:LookalikeArchitecture} depicts the Trending-Prospecting flow for training the Lookalike model. After joining the two data sources in order to obtain a User-Product feed\rotemIeee{what happens after that?}\orenIeee{do we have a different lookalike model for each advertiser or one model for all advertisers? it's a bit confusing}\naamaIeee{The threshold is being set per advertiser, but it's one model}, we derive from it the top $N$ popular products and the users that engaged with them. The top $N$ products (positive events) along with the negative random samples of Gemini impression feed for each product (negative events), are then used by \offset to train the Lookalike model. The Lookalike model is updated daily, excluding products that haven't appeared in the model for the past $10$ days \rotemIeee{there's a lot of overlapping information between this paragraph and the previous one. Maybe unify them instead of explaining things twice?}.\orenIeee{done}}\orenIeee{please have a look}After joining the pixel feed and the negative sample taken from Gemini native impression feed\naamaIeee{you mention it's the impression feed in the last sentences, I don't think it should be mentioned twice} to create the User-Product feed, it is used by \offset to update the Lookalike model. The model is updated daily, excluding products that haven't appeared in the model during the past $10$ days.
\Comment{
\begin{figure}[t]
\centering
\includegraphics[width=1.0\columnwidth]{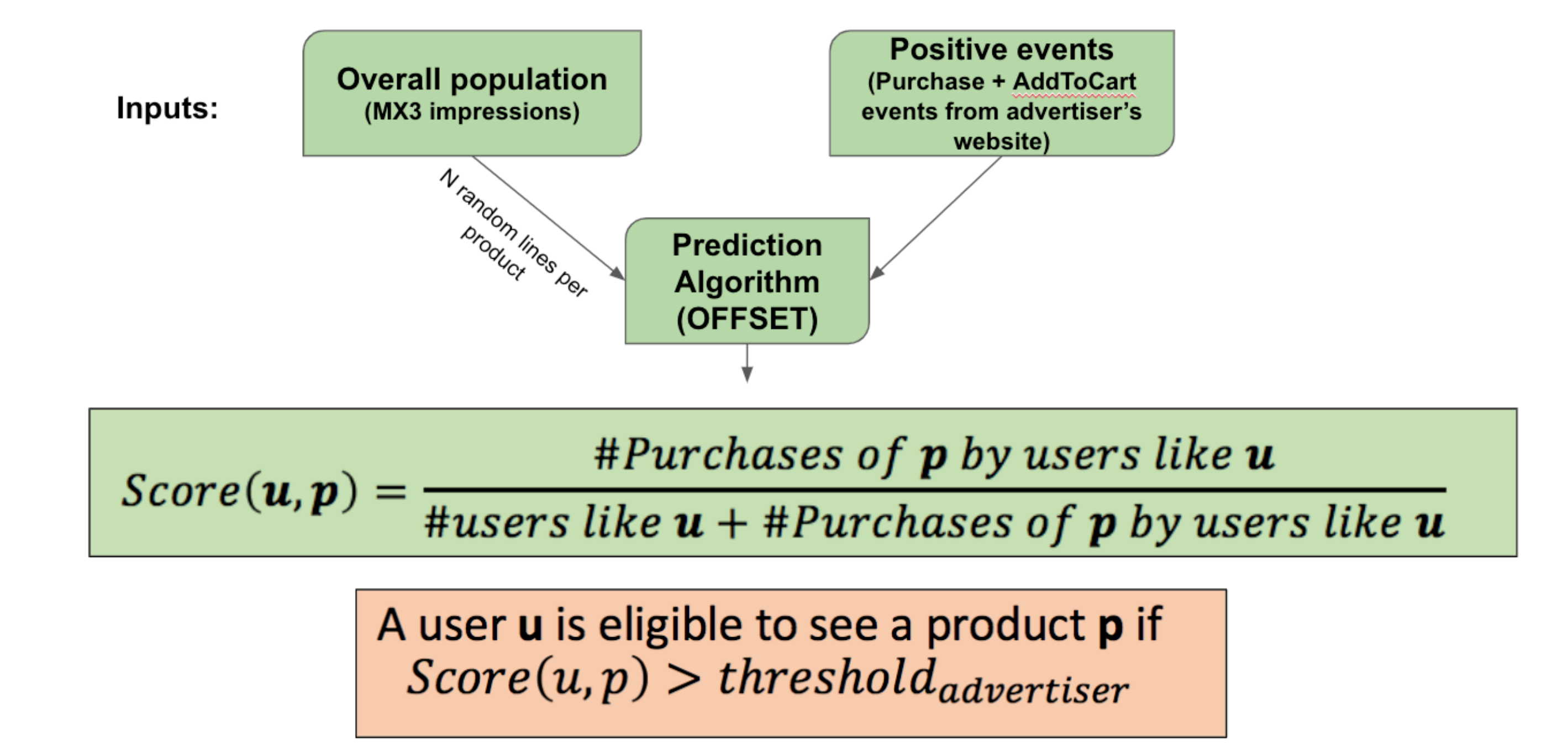}
\vspace{\mySpace}
\caption{\orenIeee{Let's talk about it. maybe share the slide so I can change it directly and save us some time} Trending-Prospecting Lookalike model training flow. \rotemIeee{Don't mention MX3 externally. Maybe rename "Overall population" to "click event", and "Positive events" to "conversion events". Also, the inputs don't match the output, the inputs are for training while the output is for serving. Also, either mark both input and output or none}}
\label{fig:LookalikeArchitecture}
\end{figure}}

Training the Lookalike\rotemIeee{\offset} model using the above inputs changes the semantics of the resulting prediction of Eq. \eqref{eq:event probability prediction}\rotemIeee{refer to the pCTR formula}\orenIeee{done}. Given a user $u$ and a product $p$, the resulting prediction (or eligibility score) will correspond to the ratio between the number of users similar to $u$\footnote{In the current setting, similar users means users with the same age and gender.} that engaged with $p$ in the advertiser pixel feed, to the number of users similar to $u$ in the general population sampled from Gemini impression feed\orenIeeeFinal{why do we need the absolute value signs to present the power of the set. we may use the same pos and neg without it}\orenIeeeFinal{done}
\begin{equation}\label{eq:Trendy-Score}
    \mathrm{S}_{u,p} \approx \frac{\#pos}{\#pos+\#neg} = \\ 
    \frac{pos(u,p)}{pos(u,p) + neg(u)}\ ,
\end{equation}
where $pos({u,p})$ is the number of users like $u$ that engaged with $p$ and $neg(u)$ is the number of users like $u$ in the $M$ negative random sample taken from the Gemini native impression feed\rotemIeee{you keep referring to this feed by different names and descriptions, you should pick one and stick with it}\orenIeee{done}. The score $\mathrm{S}_{u,p}$ captures\rotemIeee{maybe "captures" is better?} the popularity of $p$ represented by $pos(u,p)$\rotemIeee{this and the following explanation shouldn't be in brackets}\orenIeee{done} and the lookalike level of user $u$ to the users that already engaged with $p$\rotemIeee{this reads like "the score corresponds to the score"}\orenIeee{done}.
As with the Conversion-Prospecting type, since users are represented by their features, audience expansion is less restrictive than other prospecting types that require user id or bCookie\rotemIeee{mentioning both user id and bcookie is too much technical information, user id should suffice}. 

\subsubsection{Eligibility Score Threshold}
Since $\mathrm{S}_{u,p}\in [0,1]$, we set an {\it Eligibility Threshold} to determine the binary decision of $u$ being eligible/non-eligible to be presented with a DPA ad displaying $p$. Changing\rotemIeee{"Determining" might be better here} the value of the eligibility threshold gives advertisers\rotemIeee{the advertiser?}\orenIeee{done} flexibility in delivery. Advertiser can choose to decrease the threshold in order to increase delivery and expose their\rotemIeee{their}\orenIeee{done} products to a wider audience. Alternatively, advertisers can increase the threshold in order to improve performance and focus the delivery on users with higher eligibility scores that are more likely to engage with their\rotemIeee{their or the}\orenIeee{done} products. 
\Comment{
The system supports setting the eligibility threshold by defining the percentage of the population that will be considered eligible.}

The system facilitates establishing the eligibility threshold by specifying an estimated percentile of the population that will be deemed eligible. In order to do so, after each time the model is trained we randomly pick a test-sample of $R$ (e.g., $R=20K$) \naamaIeee{I might be missing something, but so far we use capital letters to mark sizes} users. Then, for each user in the test-sample we use the Lookalike model to score all available products of an advertiser, and store the user's highest eligibility score $\max_p\{\mathrm{S}_{u,p}\}$ in a histogram (per advertiser) portraying the distribution of maximal eligibility scores in the population.
\Comment{In order to do so, after each time the model is trained we generate eligibility scores $\mathrm{S}_{u,p}$\rotemIeee{which scores?} to a new test-sample of users of size $R$ (e.g., $R=20K$). For each user $u$ in the test-sample we score all available products of an advertiser \rotemIeee{you should combine this and the previous sentence}. The\rotemIeee{product with the?} highest eligibility score for each user is kept in a histogram portraying the distribution of maximal eligibility scores in the population\rotemIeee{please explain the histogram better, I couldn't follow}.}

Finally, we use the histograms to generate a threshold-percentile curve for each advertiser (i.e., score distribution). Using the threshold-percentile curves we can determine the eligibility-threshold that correspond to the advertiser's percentage setting. For example, if an advertiser wants $5\%$ of the population to be eligible to its products, we will choose $t$ from the threshold-percentile curve, so that $5\%$ of the test-sample users (and similar percentage of the general population) have an eligibility score higher than $t$ for at least one of the advertiser's products. This gives the advertiser control over the quality-quantity trade off of its Trending-Prospecting campaigns.\rotemIeee{this whole explanation is convoluted, some information is repeated twice with some new info in the middle. It's very hard to follow}\orenIeee{done}

\Comment{Given a certain threshold $t\in[0,1]$\rotemIeee{where does $t$ come from?}\orenIeee{it's a parameter at this stage} and the histograms, we calculate for each advertiser the percentage of users that will be considered as eligible, i.e., what percentage of users have a maximal score above $t$\rotemIeee{for a single product? For all products? For the advertisers products?}\orenIeee{done}. Using this mechanism we generate a threshold-percentile curve for each advertiser, from which each advertiser can set the percentage of eligible users for its Trending-Prospecting DPA ads. For example, if an advertiser sets a threshold that corresponds to $5\%$ of the population, we will choose $t$ from the threshold-percentile curve, so that $5\%$ of the test-sample users have an eligibility score higher than $t$ for some product of the advertiser. This gives the advertiser control over the quality-quantity trade off of its Trending-Prospecting campaigns}

\paragraph{Published model} After calculating the eligibility score thresholds according to the advertisers' settings, the model and the thresholds are\rotemIeee{remove "being"}\orenIeee{done} published and sent\rotemIeee{replace "as an input for" with "to"}\orenIeee{done} to the DPA Serving system (see Section \ref{subsec:DPA serving}). Due to serving limitations, the published model is bounded to $T$ (e.g., $T=3500$) products,
where product-groups with higher spend during the previous day will have more of their products included in the model.
\Comment{
while setting a minimum limit to ensure no product-group would suffer from ``starvation''}
As with Conversion-Prospecting (see Section \ref{sec:convPros}), to decrease the load on the main Gemini serving system, Trending-Prospecting products also compete in an internal preliminary auction with other prospecting products, and only the top $L$ (e.g., $L=40$) prospecting products are included in the final Gemini native auction.\orenIeee{what is the bid used for trending-prospecting?}
\vspace{-0.1cm}

\subsection{The Benefits of the new approaches}
\vspace{-0.1cm}
\Comment{
\begin{itemize}
\item{\bf Broader audience expansion} the users' eligibility to products is determined by the user and product features. Since user-identification (by user id or bCookie) or location-based targeting is more limited in scope compared to targeting based on user characteristics, the potential reach of the new types is greater than other prospecting methods.

\item{\bf Quality-quantity trade-off} both types can enhance or restrict the reach of users by modifying the performance objective. In \textit{Conversion-Prospecting} the tCPA can be adjusted according to the advertiser's needs. A lower tCPA will lead to a reduced bid and a smaller pool of eligible users and vice versa. The eligibility score threshold in \textit{Trending-Prospecting} can also be adjusted to either limit the pool of eligible users to only those with high scores and quality, or to expand the pool to include users with lower scores\rotemIeee{consider rephrasing to "a more diverse user pool"}.

\item{\bf Performance goal}\rotemIeee{The}\orenIeee{done} the \textit{Conversion-Prospecting} type leverages Gemini native data feeds \rotemIeee{consider using "logged events" instead of "feeds"} to predict the probability of a user to convert on (i.e.,\rotemIeee{use "i.e." instead of "or"}\orenIeee{done} purchase) a product.
This allows targeting only those users who are more likely to convert, and optimizing the bids to ensure that the\rotemIeee{the} target CPA is achieved.
\item{\bf Supporting new advertisers and products} the \textit{Trending-Prospecting} type uses advertisers' feeds and \textit{non-advertiser specific} negative random sample from Gemini native impression feed\rotemIeee{so not advertisers' feeds only?}\orenIeee{done}. This facilitates, the ability to accommodate new advertisers and products having no event history logged in Gemini native feeds\rotemIeee{how exactly?}\orenIeee{self explanatory }. 
\end{itemize}
}

\noindent{\bf Broader audience expansion} the users' eligibility to products is determined by the user and product features. Since user-identification (by user id or bCookie) or location-based targeting is more limited in scope compared to targeting based on user characteristics, the potential reach of the new types is greater than other prospecting methods.

\noindent{\bf Quality-quantity trade-off} both types can enhance or restrict the reach of users by modifying the performance objective. In \textit{Conversion-Prospecting} the tCPA can be adjusted according to the advertiser's needs. A lower tCPA will lead to a reduced bid and a smaller pool of eligible users and vice versa. The eligibility score threshold in \textit{Trending-Prospecting} can also be adjusted to either limit the pool of eligible users to only those with high scores and quality, or to expand the pool to include users with lower scores\rotemIeee{consider rephrasing to "a more diverse user pool"}.

\noindent{\bf Performance goal}\rotemIeee{The}\orenIeee{done} the \textit{Conversion-Prospecting} type leverages Gemini native data feeds \rotemIeee{consider using "logged events" instead of "feeds"} to predict the probability of a user to convert on (i.e.,\rotemIeee{use "i.e." instead of "or"}\orenIeee{done} purchase) a product.
This allows targeting only those users who are more likely to convert, and optimizing the bids to ensure that the\rotemIeee{the} target CPA is achieved.

\noindent{\bf Supporting new advertisers and products} the \textit{Trending-Prospecting} type uses advertisers' feeds and \textit{non-advertiser specific} negative random sample from Gemini native impression feed\rotemIeee{so not advertisers' feeds only?}\orenIeee{done}. This facilitates, the ability to accommodate new advertisers and products having no event history logged in Gemini native feeds\rotemIeee{how exactly?}\orenIeee{self explanatory }. 

\vspace{-0.1cm}
\section{Performance Evaluation}\label{sec:evaluation}
In this section we present offline and online evaluations of the new DPA Prospecting types and their impact. Since proprietary data is used, it is clear that others can not reproduce our results. This caveat is common in papers describing commercial systems and should not impair the overall contribution of the work.
\vspace{-0.1cm}
\subsection{Offline Evaluation}
In this section we present offline results for demonstrating the individual residual contributions of the user features to the Conversion-Prospecting model (see Section \ref{sec: conv-pros model}), and the Trending-Prospecting Lookalike model (see Section \ref{sec: lookalike model}) accuracies. We use the \textit{wrapper forward selection method} \cite{featureSelection2003} - an iterative method where in each iteration we add one new feature to the model which shows the best performance among all other features when compared to the performance of the model resulting from the previous iteration. \rotemIeee{you should explain why these methods were not compared to anything else but themselves}
As baselines we started with models operating with product features only. 
In this test we trained all models involved from ``scratch''\footnote{Training from ``scratch'' means initializing the model's parameters in a \textit{lazy} fashion using  a Normal distribution with zero mean and small variance.} over six weeks worth of data, and measured the performance
 on the logged data of the following week. 
 The Conversion-Prospecting models were updated every 6 hours on a\orenNew{switched to weekly numbers}\orenNew{agreed} weekly average of a ${\sim} 7$K conversions and ${\sim} 1.4$M clicks.
 The Trending-Prospecting Lookalike models were updated daily using $2$K negative samples for every product\footnote{A daily average of ${\sim} 20$M impressions, since the model is updated daily with $10K$ top popular products.}, and a daily average of ${\sim} 165$K positive events (\textit{purchases} and \textit{add-to-cart} events).
 Both models use \offset's internal \textit{adaptive online hyper-parameter tuning mechanism} \cite{aharon2017adaptive}, to tune \offset algorithm hyper-parameters (see Appendix \ref{sec: offset}).
 
To evaluate the user features' contributions we use the following performance metrics.
\vspace{-0.05cm}
\paragraph{Logistic loss (LogLoss)}
\[
\sum_{(u,p,y)\in \mathcal{T}} \! \! -y \log
\left(\mathrm{Q}_{u,p}\right)-(1-y)\log\left(1-\mathrm{Q}_{u,p}\right)\ ,
\]
where ${\mathcal{T}}$ is a test set, $\mathrm{Q}_{u,p}$ is the model conversion-given-click probability prediction (or eligibility score)\rotemIeee{you already defined pCONV, no need to do that again}\orenIeee{done}, and $y \in \{0,1\}$ is the positive event indicator.
We note that the LogLoss metric (where a lower value is better) is used to train the models and to tune \offset algorithm hyper-parameters (see Appendix \ref{sec: offset}).
\vspace{-0.2cm}
\paragraph{Area-under ROC curve (AUC)} The AUC (where a higher\rotemIeee{a higher value}\orenIeee{done} value is better) is equivalent to the probability that the pairwise ranking of the predictions (or scores) of two random events (one positive and one negative), is correct \cite{fawcett2006introduction}.
\vspace{-0.1cm}
\begin{table}[ht]
  \centering
  \caption{Conversion model \textit{forward-selection} results.}
  \vspace{-0.2cm}
  \label{tab:conv pros offline}
  \begin{tabular}{|c|c|c|c|}
    \hline
    \# Feature & User feature & AUC lift & Logloss lift \\
    \hline
    1\ (1)& ctr-campaign-top & $2.97\%$ & $13.52\%$ \\
    \hline
    2\ (1+2)& dpa-type-experiment-id & $3.36\%$ & $17.49\%$ \\
    \hline
    3\ (1+2+3)& page-section & $3.60\%$ & $18.87\%$ \\
    \hline
  \end{tabular}
  \end{table}
\vspace{-0.1cm}
\subsubsection{Conversion-Prospecting}
The individual contributions of the user features to the Conversion-Prospecting model accuracy in terms of LogLoss and AUC lifts\footnote{Lifts are calculated for any positive metric $M>0$ by $(M_{model}/M_{baseline}-1)\cdot 100$ for ``higher is better'' metrics and by $(M_{baseline}/M_{model}-1)\cdot 100$ for ``lower is better'' metrics.} over the product-features-only baseline model are presented in Table \ref{tab:conv pros offline} \rotemIeee{you should note that lift means improved metric, because in LogLoss a lower value is better}\orenIeee{done in a footnote}.
It is interesting to observe that a click related feature provides the ``strongest'' indication of users' conversion patterns \rotemIeee{Do you mean ctr-campaign-top? If you didn't try a different sequence of feature insertion and saw that you get the highest initial lift with ctr-campaign-top then can't really say that}. In addition to the features outlined in Table \ref{tab:conv pros offline}, we also tested other features that did not lead to an improvement in the model's accuracy. For instance, the \textit{device-type} feature provided redundant information that seemed to be already captured by the \textit{page-section} feature. As a result, no enhancements were observed in the model's AUC and Logloss metrics.
\vspace{-0.1cm}
\begin{table}[ht]
  \centering
  \caption{Lookalike model \textit{forward-selection} results.}
  \vspace{0.1cm}
  \label{tab:trendy offline}
  \vspace{\mySpace}
  \begin{tabular}{|c|c|c|c|}
    \hline
    \# Feature & Feature & AUC lift & Logloss lift \\
    \hline
    1\ (1) & age & $5.55\%$ & $1.94\%$ \\
    \hline
    2\ (1+2) & gender & $10.99\%$ & $6.09\%$ \\
    \hline
  \end{tabular}
\end{table}
\vspace{-0.1cm}
\subsubsection{Trending Prospecting}
\naamaNew{Copied from conv-pros section}
Similarly to the process mentioned above, the individual contributions of the user features to the Trending-Prospecting Lookalike model accuracy in terms of LogLoss and AUC lifts over the product-features-only baseline model are presented in Table \ref{tab:trendy offline}. Interestingly, the combined lifts of age and gender are much higher than their individual lifts.
\naamaNew{Copied from conv-pros. Do we need to think of another fact regard trendy test?}

\subsection{Online Evaluation} \label{subsec: online res}
Recall that the new DPA types have two main goals: (1) increase DPA audience; and (2) maintain predefined performance goals.
\subsubsection{Conversion-Prospecting}
To assess whether these two goals are achieved, we launched two online production-like buckets, serving $10\%$ of Gemini native traffic each, for a period of two weeks; a test bucket operating with Conversion-Prospecting and a control bucket operating without it.

\begin{figure}[t]
\centering
\includegraphics[width=0.8\columnwidth]{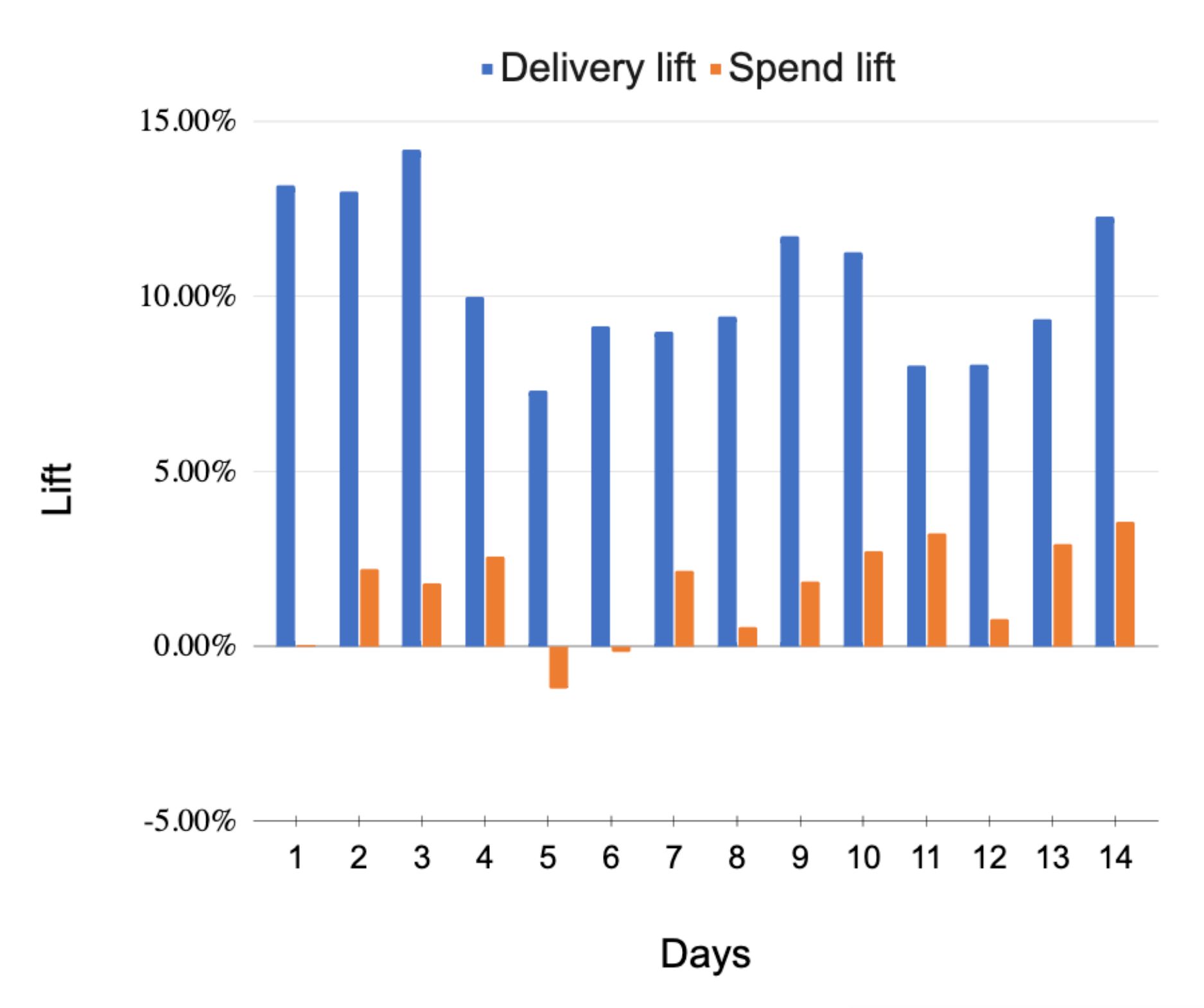}
\vspace{\mySpace}
\caption{\orenNew{Can you add the days ticks?}\naamaNew{done - it became a bit bigger}\orenNew{you can control the proportion and please make all text bigger} Conversion-Prospecting daily spend and delivery lifts.\orenIeee{change "impressions" to "delivery"}\naamaIeee{done}}
\label{fig: conv daily}
\end{figure}

\paragraph{DPA audience}\label{subsec: dpa audience conv pros}
To evaluate how Conversion-Prospecting affects the DPA audience we measure the daily spend (or revenue) and delivery (or number of impressions) of the two buckets over a period of two weeks. The daily spend and delivery lifts of the test bucket over the control buckt\rotemIeee{of the experiment bucket over the baseline}\orenIeee{done} are plotted in Fig. \ref{fig: conv daily} and reveal positive lifts during most days. In particular,\rotemIeee{"In particular" doesn't make sense in this context} we measure $1.58\%$ average DPA spend lift, and more than $10\%$ average DPA delivery lift over the whole test period\rotemIeee{over the whole period of time?}\orenIeee{done}.\oren{what about CPM? it's not necessarily higher?}

\paragraph{Performance goal}
To evaluate the Conversion-Prospecting performance goal, we measure its \textit{cost-per-action} (CPA)\rotemIeee{you already defined the CPA acronym}\orenIeee{i know} for each advertiser, which is defined as\rotemIeee{no need to repleat the formula in words, just say "defined in formula x"} the advertiser's total spend divided by the number of the advertiser's post-click conversions\rotemIeee{it's cumbersome to read a formula with "advr.", just use $a$ to indicate an advertiser, eg $CPA_a$}
\naamaNew{remved conv_pros from the def}
\begin{equation*}
    CPA(advr.) = \frac{spend(advr.)}{\# conversions(advr.)}\ .
\end{equation*}
\Comment{
\begin{equation*}
    CPA_{conv.-pros.}(advr.) = \frac{spend(advr.)}{\# conversions(advr.)}\ .
\end{equation*}}
Next, we define the advertiser's ``happiness'', stating that an advertiser is satisfied (or ``happy'') if its CPA divided by its target CPA (tCPA) is smaller than $(1+error)$, for allowing an error margin
\Comment{
Then, we measure the advertiser's ``happiness''\rotemIeee{consider rephrasing to "Then we define a measure we call \it{advertiser happiness}, where" and then just show the formula without explaining it with words}, saying that an advertiser is satisfied (or ``happy'')
if Conversion-Prospecting CPA divided by the target CPA (tCPA) is smaller than $(1+error)$, allowing for an error margin}
\naamaNew{Added CPA(advr.)}
\begin{equation} \label{eq:e-happy}
advr.\ is\ happy \iff \frac{CPA(advr.)}{tCPA(advr.)} \leq 1+ error\ .
\end{equation}
\Comment{
\begin{equation} \label{eq:e-happy}
Advr.\ is\ Happy \iff \frac{CPA}{tCPA} \leq 1+ error\ .
\end{equation}
}
As mention in Section \ref{sec:tCPA}, we use the Retargeting DPA type as a reference point, and set the Conversion-Prospecting performance goal to ($1.5 \cdot \mathrm{CPA_{retarg.}}$). 
\rotemIeee{which model? where did you determine that it's accurate?}\orenIeee{done}
\rotemIeee{if you define here how you reached 1+error=1.515 then there's no need for the equation below below the following equation. Either way you should explain how you reached this number}
Setting a modest error margin of $error=1\%$ and substituting $\text{tCPA}(advr.)=1.5 \cdot \mathrm{CPA_{retarg.}(advr.)}$ in Eq. \eqref{eq:e-happy}, the Conversion-Prospecting advertisers' overall ``happiness'' is given by
\begin{equation}\label{eq:happiness}
    Happiness = \frac{\sum_{advr.\ is\ happy}\ spend(advr.)}{\sum_{advr.}\ spend(advr.)}\ ,
\end{equation}
where
\begin{equation*}
advr.\ is\ happy \iff \frac{CPA_{conv.-pros.}(advr.)}{CPA_{retarg.}(advr.)} \leq 1.515\ .
\end{equation*}
We measure the CPA for Retargeting and Conversion-Prospecting DPA types for all advertisers that have at least $10$ conversions during a period of two weeks that overlaps with the period we have measured the DPA audience lifts\rotemIeee{rephrase this sentence, it's difficult to understand}. Summing-up the overall advertisers' ``happiness'' according to Eq. \eqref{eq:happiness}\rotemIeee{this equation is already a sum, you don't sum the equation}, we measure $84\%$ ``happiness'' of the total spend. We emphasize that such ``happiness'' level is considered quite high\rotemIeee{"in comparison to other conversion models launched in Gemini native" and remove the rest of the sentence}\orenIeee{done},
in comparison to other conversion models launched in Gemini native.

\subsubsection{Trending-Prospecting}
\paragraph{DPA Audience} \label{subsec: dpa audience trendy}
To evaluate how Trending-Prospecting affects the DPA audience, we launched two production-like $5\%$ buckets, serving Gemini native traffic, for a period of two weeks; A control bucket operating without Trending-Prospecting and a test bucket operating with Trending-Prospecting \rotemIeee{there will be questions why the buckets are not the same size. Maybe just say "we similarly launched a a 2\% production-like test bucket operating with Trending-Prospecting and compared it to the baseline bucket without rending-Prospecting"}\orenIeee{done}.

\begin{figure}[t]
\centering
\includegraphics[width=0.8\columnwidth]{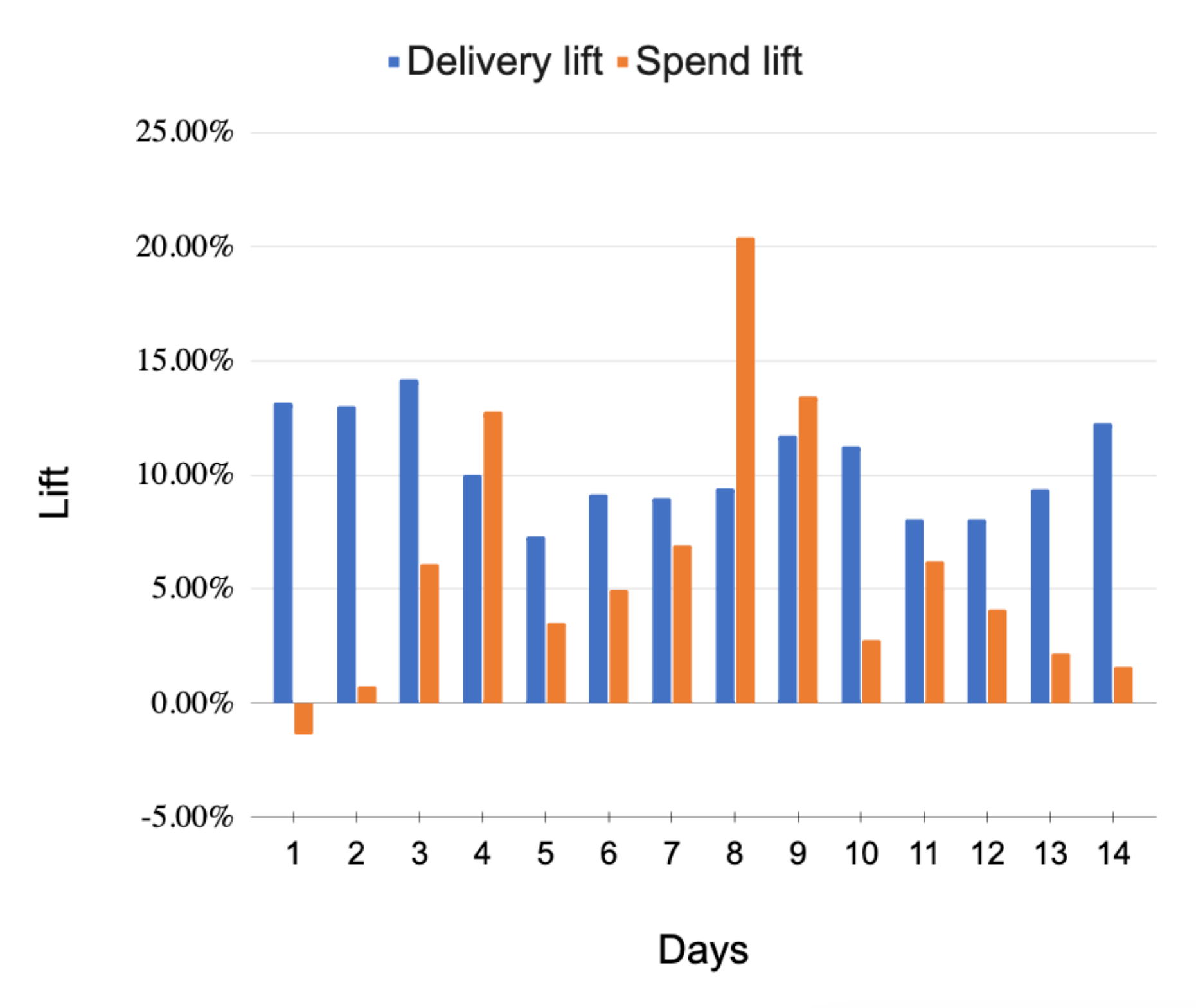}
\vspace{\mySpace}
\caption{Trending-Prospecting daily spend and delivery lifts.\orenIeee{change "impressions" to "delivery"}\naamaIeee{done}}
\label{fig: trendy daily}
\end{figure}

The daily spend (or revenue) and delivery (or number of impressions) lifts are plotted in Fig. \ref{fig: trendy daily} and reveal positive lifts\rotemIeee{of the experiment over the baseline}\orenIeee{done} of the test bucket over the control bucket during most days. In particular,\rotemIeee{not the correct phrase in this context}, we measure $6.33\%$ average DPA spend lift, and almost $13\%$ average DPA delivery lift\rotemIeee{over the whole period?}\orenIeee{done} over the whole test period.

\Comment{The DPA revenue lift of the production-like bucket when compared to the control bucket operating without Trending-Prospecting during a period of two weeks, reveals a $6.33\%$ DPA revenue lift, and an almost $13\%$ delivery lift (or the number of DPA impressions). The daily lifts are plotted in Fig. \ref{fig: trendy daily}.
Moreover, the bucket revenue and impressions on all of the traffic are flat-positive as expected (since DPA share is about $5\%$ of the overall traffic).}
\vspace{-0.05cm}

\paragraph{Performance goal} \label{subsec: dpa performance trendy}
To verify if a Trending-Prospecting advertiser is ``happy'', as defined in Eq. \eqref{eq:e-happy}, we measure each advertiser's CPA in both\rotemIeee{test and control}\orenIeee{done}, test and control buckets. By setting the control bucket's CPA as the target CPA (tCPA), and allowing an error margin $error=10\%$, the Trending-Prospecting advertiser's ``happiness'' of Eq. \eqref{eq:e-happy} is defined as
\naamaNew{added here advr. as well}
\begin{equation*}
advr.\ is\ happy \iff \frac{CPA_{test}(advr.)}{CPA_{control}(advr.)} \leq 1.1\ .
\end{equation*}
\Comment{
\begin{equation*}
Advr.\ is\ Happy \iff \frac{CPA_{test}}{CPA_{control}} \leq 1.1\ .
\end{equation*}}
\Comment{To evaluate if Trending-Prospecting advertiser is ``happy'', as defined in Eq. \eqref{eq:e-happy}, we measure if the advertiser's CPA matches the performance goal as defined in Section \ref{sec: trendy}. In particular, we measure the advertiser's CPA in both\rotemIeee{test and control}\orenIeee{done}, test and control buckets, and set \rotemIeee{set}\orenIeee{done} the control bucket's CPA as the target CPA (tCPA). By allowing a $error=10\%$ error margin, the Trending-Prospecting advertiser's ``happiness'' criterion reduces to \rotemIeee{how did you reach 1.1?}
\Comment{To evaluate if an advertiser is happy, as defined in Eq. \eqref{eq:e-happy}, we measure if the advertiser's CPA matches our performance goal as defined in Section \ref{sec: trendy}. For that we measure the advertiser's CPA in the base bucket (with trendy-prospecting) compare to the test bucket (without trendy-prospecting) CPA, setting the test bucket CPA as the target. By allowing a $10\%$ error margin, trendy-prospecting happiness define as follow:}
\naamaNew{added here advr. as well}
\begin{equation*}
Advr.\ is\ Happy \iff \frac{CPA_{test}(advr.)}{CPA_{control}(advr.)} \leq 1.1\ .
\end{equation*}
\Comment{
\begin{equation*}
Advr.\ is\ Happy \iff \frac{CPA_{test}}{CPA_{control}} \leq 1.1\ .
\end{equation*}}}
We measure the CPA for all advertisers that have at least $10$ conversions in each bucket during a period of two weeks that overlaps with the period used to measure the DPA audience lifts\rotemIeee{rephrase}. Summing-up\rotemIeee{again, the equation is already a sum, you don't sum over it} the overall advertisers' ``happiness'' according to Eq. \eqref{eq:happiness} and allowing an error of $10\%$, we measure $88\%$ ``happiness'' of the total spend. As mentioned earlier, such ``happiness'' level is considered quite high.  

\subsubsection{Impact}
\begin{figure}[t]
\centering
\includegraphics[width=0.85\columnwidth]{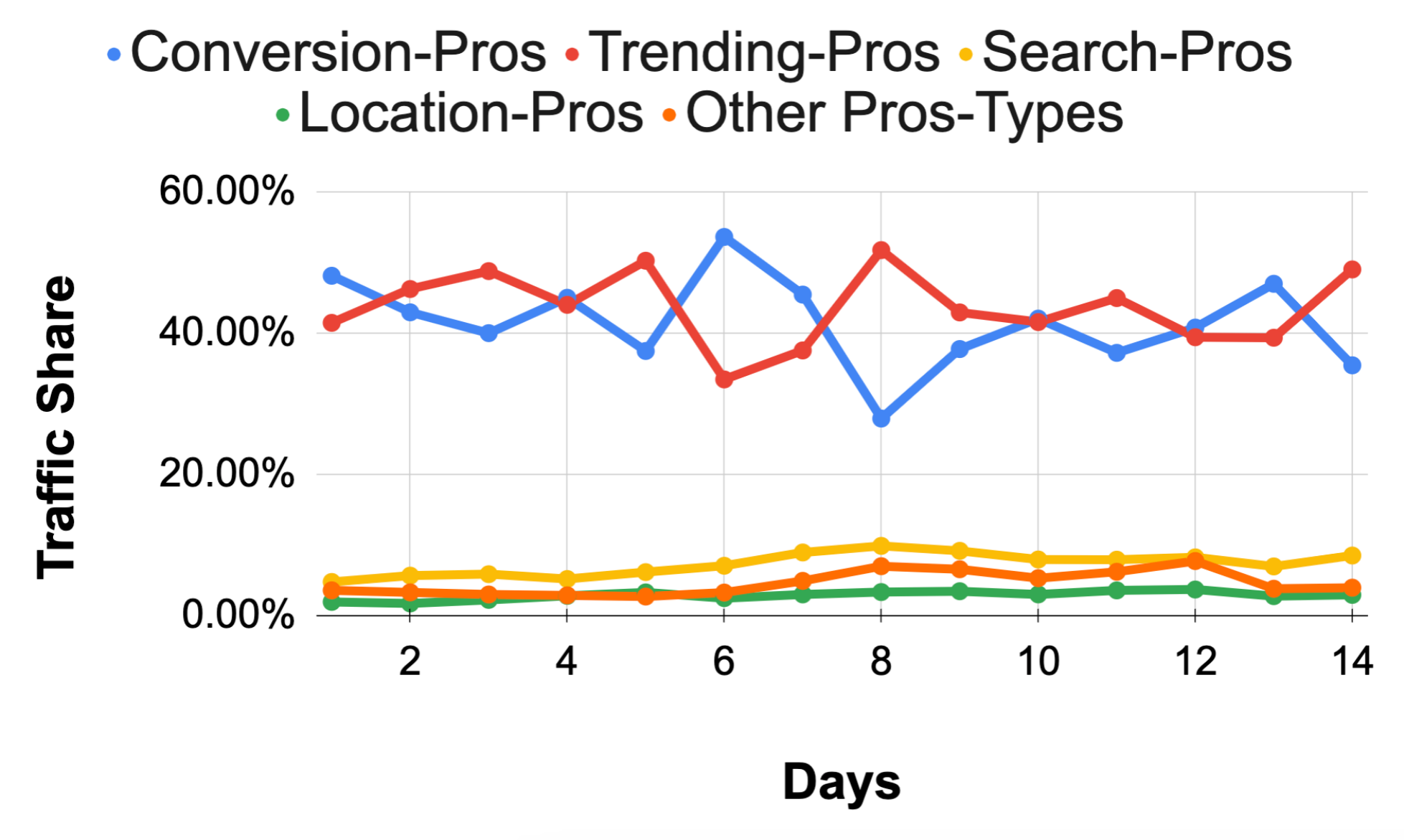}
\vspace{-0.1cm}
\caption{The daily relative traffic share of Prospecting types.\rotemIeee{mention that the blue and red lines are the new additions}\orenIeee{the legend should be improved}\naamaIeee{Is it better?}}
\label{fig: pros impr share}
\end{figure}

Soon after launching the new DPA types into production\rotemIeee{to 100\%? if they went through the whole ramp up process, why do you report numbers for different bucket sizes?}, we analyzed the traffic share\orenIeee{spend or impressions?}\naamaIeee{impressions} of various DPA Prospecting types during a period of two weeks, focusing on advertisers who share their conversion data with Yahoo. As is shown in Fig. \ref{fig: pros impr share}, the new types account for a much larger share of traffic compared to other Prospecting types, with each of the two new types serving over $40\%$ of the total Prospecting traffic on average\rotemIeee{did the new types replace other prospecting types, or is this in addition? Because if they replaced them then you should show that they perform better, otherwise you should mention that they didn't}.\orenIeee{what percent of the new types traffic was new DPA traffic and what percentage was cannibalized?}\naamaIeee{We can only check this by using the isolated budget system. I can make an estimation in case we still have the data on the bucket without trendy and see the drop in the other prospecting type spend but it's a lot of work so let me know if it's crucial. Also the delivery and spend graphs show the increase in DPA overall.}

\section{Conclusion and future work}\label{sec:Concluding remarks}
In this work we have considered Gemini native \textit{dynamic-product-ad} (DPA)\rotemIeee{you already defined the DPA acronym, no need to do that again}\oren{done}\rotemIeee{consider using a different name than "product", it's confusing}\orenIeee{done}, which is one of Yahoo's fastest growing businesses. This product uses advertisers'\rotemIeee{product} catalogs to generate ads that participate in Gemini native auctions. After describing the DPA system, we introduced two new Prospecting types that are aimed to expand DPA reach while maintaining a predefined performance goal. Since the new DPA types establish the eligibility of users to catalog products based on users' and products' features, they are not limited to a bounded set of users as other DPA Prospecting types that require user id, bCookie, or locations.

The Conversion-Prospecting type applies a conversion prediction model trained on Gemini native feeds to determine the eligibility of an incoming user to catalog products. In particular, a user is eligible to a product only if the expected \textit{cost-per-action} (CPA)\rotemIeee{already defined acronym} meets the advertisers' expected goal. Online performance evaluation reveals that the new DPA type increases\orenIeeeFinal{increases}\orenIeeeFinal{done} DPA audience by more than $10\%$ and DPA revenue by $1.58\%$, while maintaining $84\%$ of its spend within the advertisers' performance goal.

The Trending-Prospecting supports new advertisers with no event history in Gemini native feeds. We use advertisers' feed to train a lookalike model to assess the eligibility of an incoming user to catalog products, depending on the product popularity and the similarity level of the user to users already engaged with the product. Online performance evaluation reveals that the new Trending-Prospecting type increases\orenIeeeFinal{increases}\orenIeeeFinal{done} DPA audience by almost $13\%$ and DPA revenue by $6.33\%$, while maintaining $88\%$ of its spend within the advertisers' assumed performance goal. 

The two new Prospecting types are now fully deployed in production serving all Gemini native traffic. Moreover, measurements made after the full deployment reveal that the new Prospecting types traffic share exceeds $80\%$ of all DPA Prospecting traffic of advertisers that share their conversion information with Yahoo.

Future work includes a transition to a \textit{Broad-Audience framework}, where bid modifiers are automatically assigned to all DPA types. This is probably the next generation of DPA, since advertisers will be oblivious to the various DPA types used and still obtain the best performance for their spend.

\appendix
\subsection{{OFFSET}: Yahoo Event-Prediction Algorithm}\label{sec: offset}
\rotemIeee{It's not clear that this is the appendix. Is it supposed to look like this?}The Gemini native models are powered by \offset (One pass Factorization of Feature Sets), which is an ad event prediction algorithm that utilizes a feature enhanced \textit{Collaborative-Filtering} (CF) \cite{aharon2013off}\cite{aharon2017adaptive}\cite{aharon2019soft}\cite{arian2019feature}\cite{kaplan2021dynamic}. 
According to \offset, the \textit{predicted event probability} (pET) of a user $u$ and an ad $a$ is
\begin{equation}\label{eq:event probability prediction}
    \mathrm{pET}_{u,a} = \sigma(s_{u,a})\in [0,1]\ ,
\end{equation}
where $\sigma(x)=\left(1+e^{-x}\right)^{-1}$ is the \textit{Sigmoid} function, and 
\begin{equation}\label{eq: score}
	s_{u,a}=b+\nu_{u}^T \nu_{a}\ ,
\end{equation}
$\nu_{u},\ \nu_{a}\in \R^D$ denote the user and ad \textit{latent factor} (LF) vectors respectively, and $b\in \R$ denotes the model bias. The product $\nu_{u}^T \nu_{a}$ indicates the tendency of user $u$ towards ad $a$, where a higher score implies a higher predicted event probability (pET). It is noted that the model parameters $\Theta$, which are used to construct ${\nu_{u},\nu_{a}}$, are learned from the data.

\offset is utilized to power several of Gemini's native models, including the Click model which predicts a click event (pCTR), the Conversion model which predicts a conversion-given-click event (pCONV), and the Ad close model which predicts an ad close event (pCLOSE) \cite{silberstein2020ad}.

\begin{figure}[t]
\centering
\includegraphics[width=0.8\columnwidth]{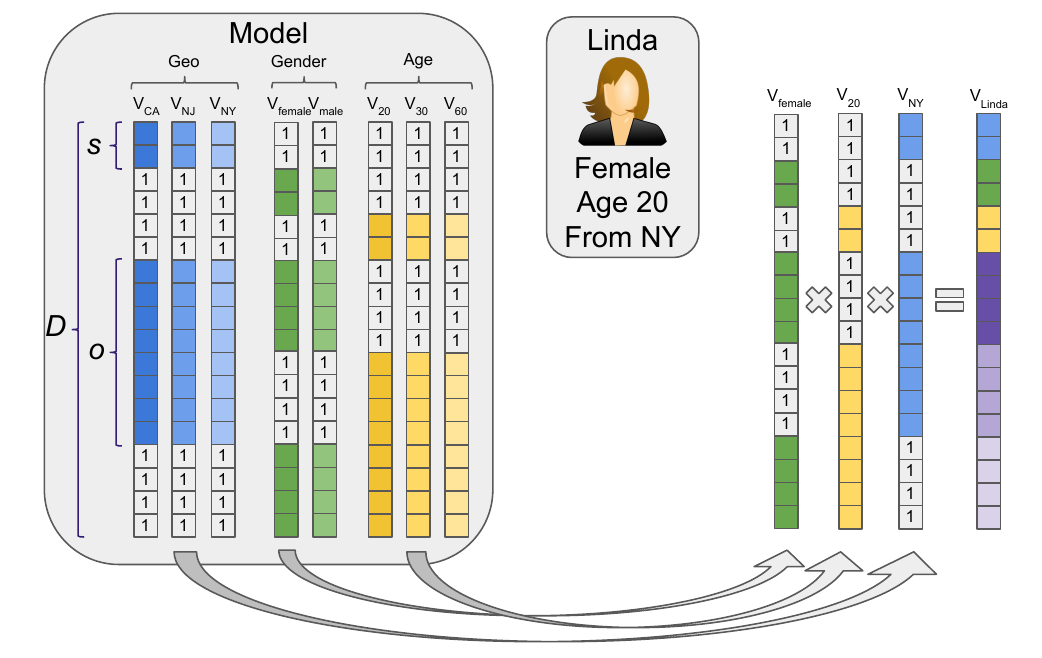}
\vspace{-0.2cm}
\caption{An example of a user LF vector for $o=4,\ s=2$ and $K=3$ features. The user vectors of dimension $d=10$, are obtained from the model, and entry-wise multiplied to produce the final user vector of dimension $D=18$ on the right.}
\Comment{
\caption{An example of constructing a user latent factor vector for $o=4,\ s=2$ and $K=3$ features (i.e. age, gender, and geo). The user feature value vectors of dimension $d=10$, are obtained from the model, filled with 1s, and entry-wise multiplied to produce the final user vector of dimension $D=18$ on the right.\oren{new: did you chatGBT the caption as well?}\naama{done}}
}
\label{fig: user vector construction}
\end{figure}

To address the problem of data sparsity (as ad events like clicks, conversions, or closes are infrequent), the ad and user feature vectors are constructed using their features. The ad feature vector is created by summing up their individual feature vectors (such as ad id, campaign id, advertiser id, ad categories, etc.) all in dimension $D$. Meanwhile, the interaction between the various $d$-dimension user feature LF vectors to form the user's $D$-dimension LF vector is more complex to account for non-linear dependencies between feature pairs. 
Accordingly, the user vector is generated based on learned vectors $\{v_k\}$ for $K$ features (e.g., gender, age, device type, geo), each of dimension $d$.\orenIeeeFinal{rephrase the last part}\orenIeeeFinal{done}
Each pair of user features is allocated with $o$ entries, while\orenIeeeFinal{"while" (?)}\orenIeeeFinal{done} $s$ entries are designated for each feature vector individually. Thus, the dimension of a single feature value vector is $d=(K-1)\cdot o + s$, and the dimension of the combined user vector is $D=\binom{K}{2} \cdot o + K\cdot s$, where $\binom{K}{2}$ represents the number of combinations of $K$ items taken 2 at a time. This construction is illustrated in Fig. \ref{fig: user vector construction}. 

Compared to traditional Collaborative Filtering (CF), the model only requires $O(K)$ LF vectors, one for each feature value (e.g., two for gender: female and male) instead of millions of unique user LF vectors.

To determine the model parameters $\Theta$, \offset minimizes the Logistic loss (or LogLoss) of the training data set $\mathcal{T}$ (past negative and positive events) using a one-pass \textit{online gradient descent} (OGD)
\vspace{-0.1cm}
\begin{equation*}
\argmin_{\Theta}\!\!\!\! \sum_{(u,a,y)\in \mathcal{T}} \mathcal{L}(u,a,y)\ ,
\end{equation*}
\vspace{-0.1cm}
where $\mathcal{L}(u,a,y)$ equals
\Comment{\begin{multline}
\mathcal{L}(u,a,y)=\\-(1-y)\log\left(1-\mathrm{pEVENT}_{u,a}\right)-y \log \left(\mathrm{pEVENT}_{u,a}\right)+\frac{\lambda}{2}\sum_{\theta\in\Theta}\theta^2\ ,
\end{multline}}
\vspace{-0.1cm}
\[
-(1-y)\log\left(1-\mathrm{pET}_{u,a}\right)-y \log \left(\mathrm{pET}_{u,a}\right)+\frac{\lambda}{2} \|\Theta\|_2^2\ ,
\]
\vspace{-0.1cm}
$y \in {0,1}$ is the label (or indicator) for the event involving user $u$ and ad $a$, and $\lambda$ indicates the $L2$ regularization factor. The step sizes for OGD are determined using a variant of the AdaGrad algorithm \cite{duchi2011adaptive} as described in \cite{aharon2017adaptive}.

The \offset algorithm is based on incremental training, updating its model parameters with every batch of new training events, such as every 15 minutes for the click model or 4 hours for the conversion model. It includes an \textit{adaptive online hyper-parameter tuning mechanism} \cite{aharon2017adaptive} that uses the parallel map-reduce architecture of the Gemini platform to adapt to changing marketplace\orenIeeeFinal{marketplace}\orenIeeeFinal{done} conditions, like trends and temporal effects. This mechanism tunes \offset's hyper-parameters, like the OGD initial step size and AdaGrad parameters. In addition, it also supports dynamic allocation of user feature LF vector sizes (implementing an internal automatic ``feature selection'' mechanism) \cite{kaplan2021dynamic}.\naama{new, please revisit}\oren{done} Other elements of \offset, such as the similarity weights for implementing ``soft'' recency and frequency regulations
\cite{aharon2019soft}, are not included here for brevity.
\Comment{
Other components of \offset, like the similarity weights for applying "soft" recency and frequency rules (how recent and how often a user has seen the same ad or campaign) \cite{aharon2019soft}, and a mechanism that chooses dynamically the length of the model's vector and the allocation of each independent and overlaps features  \cite{kaplan2021dynamic} are not included here for brevity.
}
\oren{new: I would add Rina's paper briefly here by saying that the hyper-parameter tuning is also powering the DVA and DVL. one sentence and say for more details see \cite{kaplan2021dynamic}}\naama{done}
\vspace{-0.1cm}
\Comment{
\subsection{Weighted multi-value feature type}\label{sec:offset wmv feature}
\offset algorithm most generic feature type is the \textit{weighted multi-value} (WMV) feature type. To support this feature type, the model includes a $d$-dimension LF vector for each of the $m$ feature values encountered so far. Hence, the $d$-dimension vector of this WMV feature for user $u$ \orenIeee{fix the footnote} is\footnote{The definition is valid for ad entities as well, however, the feature values LF vectors and final vector in this case are of dimension $D$.} 
\[
v=\frac{1}{\sqrt{n}}\sum_{i=1}^n w_{\ell_i}\ v_{\ell_i}\ ,
\]
where $\{\ell_i\}$ and $\{w_{\ell_i}\}$ are the $n\le m$ feature values and accompanied weights associated with user $u$, and $\{v_{\ell_i}\}$ are the model LF vectors associated with values $\{\ell_i\}$. It is noted that the weights are given as part of the users' data and are not model parameters that are needed to be learned. Each time \offset encounters a new value for the feature, it assigns a random Gaussian vector to it with zero mean and covariance matrix $\eta\cdot I_d$, where $0\le \eta\ll 1$ and $I_d$ is the identity matrix.
It is noted that simpler feature types such as \textit{categorical features} (e.g., gender and age), or non-weighted multi-value features (e.g., all weights equal 1), may be seen as special cases of the WMV feature type.
}

\Comment{
The \offset algorithm primarily uses the \textit{weighted multi-value feature} type. This means that for each of the $m$ feature values seen so far, the model has a $d$-dimensional LF vector associated with it. The resulting $d$-dimensional vector of the weighted multi-value feature for user $u$ is 
\[
v=\frac{1}{\sqrt{n}}\sum_{i=1}^n w_{\ell_i}\ v_{\ell_i}\ ,
\]
where ${\ell_i}$ are the $n$ feature values associated with user $u$, and ${w_{\ell_i}}$ are the weights accompanying these values. The model LF vectors associated with the values ${\ell_i}$ are ${v_{\ell_i}}$.
The same definition is also valid for ad entities, with the feature values' LF vectors and resulting vector having a dimension of $D$.

It's important to note that the weights are already given as part of the user data and do not need to be learned by the model. Whenever \offset encounters a new value for a feature, it assigns it a random Gaussian vector with zero mean and covariance matrix $\eta\cdot I_d$, where $0\le \eta\ll 1$ and $I_d$ is the identity matrix. This represents one of the ``cold-star'' strategies. Other, more advanced strategies are not covered in this work. It's also worth noting that simpler feature types, such as categorical features (e.g. age and gender) or non-weighted multi-value features (e.g. user category feature with all weights equal to 1), are just special cases of the weighted multi-value feature type.}

\Comment{
\vspace{-0.2cm}
\appendix
\section{{OFFSET}: Event-Prediction Algorithm}\label{sec: offset}
The Gemini native models are powered by \offset (\textit{One pass Factorization of Feature SETs}), which is an event prediction algorithm that utilizes a feature enhanced \textit{Collaborative-Filtering} (CF) approach \cite{aharon2013off,aharon2017adaptive,aharon2019soft}\cite{arian2019feature}\cite{kaplan2021dynamic}\cite{kaplan2021unbiased}. 
According to \offset, the \textit{predicted event probability} (pET) of a user $u$ and an ad $a$ is
\begin{equation}\label{eq:event probability prediction}
\mathrm{pET}_{u,a} = \sigma(b+\nu_{u}^T \nu_{a})\in [0,1]\ ,
\end{equation}
where $\sigma(x)=\left(1+e^{-x}\right)^{-1}$ is the \textit{Logistic Sigmoid} function, 
\Comment{\begin{equation*}\label{eq: score}
	s_{u,a}=b+\nu_{u}^T \nu_{a}\ ,
\end{equation*}}
$\nu_{u},\ \nu_{a}\in \R^D$ denote the user and ad \textit{latent factor} (LF) column vectors respectively, and $b\in \R$ denotes the model bias. The product $\nu_{u}^T \nu_{a}$ indicates the tendency of user $u$ towards ad $a$, where a higher score implies a higher predicted event probability (pET).  It is noted that the model parameters $\Theta$, which are used to construct ${\nu_{u},\nu_{a}}$, are learned from the data as will be explained later.

\Comment{\offset is utilized to power several of Gemini's native models, including the Click model which predicts a click event (pCTR), the Conversion model which predicts a conversion-given-click event (pCONV), and the Ad close model which predicts an ad close event (pCLOSE) \cite{silberstein2020ad}.}

To address the problem of data sparsity (as ad events like clicks, or conversions are infrequent), the ad and user feature vectors are constructed using their features. The ad feature vector is created by summing up their individual feature vectors (such as ad id, campaign id, advertiser id, ad categories, etc.) all in dimension $D$. Meanwhile, the interaction between the various $d$-dimension user feature LF vectors to form the user's $D$-dimension LF vector is more complex to account for non-linear dependencies between feature pairs. The user vectors are created using their $K$ features' learned vectors $v_k$ (such as gender, age, device type, geo, etc.) which each have dimension $d$. Each pair of user features is allocated $o$ entries, and $s$ entries are designated for each feature vector individually. Thus, the dimension of a single feature value vector is $d=(K-1)\cdot o + s$, and the dimension of the combined user vector is $D=\binom{K}{2} \cdot o + K\cdot s$, where $\binom{K}{2}$ represents the number of combinations of $K$ items taken 2 at a time. For more details see \cite{kaplan2021dynamic}.
Compared to traditional CF algorithms, the model only requires $O(K)$ user side LF vectors, one for each feature value (e.g., two for gender: female and male) instead of millions of unique user LF vectors. 

To determine the model parameters $\Theta$, \offset minimizes the Logistic loss (or LogLoss) of the training data set $\mathcal{T}$ (past negative and positive events) using a one-pass \textit{online gradient descent} (OGD)
\begin{equation*}
\argmin_{\Theta}\!\!\!\! \sum_{(u,a,y)\in \mathcal{T}} \mathcal{L}(u,a,y)\ ,
\end{equation*}
where $\mathcal{L}(u,a,y)$ equals
\Comment{\begin{multline}
\mathcal{L}(u,a,y)=\\-(1-y)\log\left(1-\mathrm{pEVENT}_{u,a}\right)-y \log \left(\mathrm{pEVENT}_{u,a}\right)+\frac{\lambda}{2}\sum_{\theta\in\Theta}\theta^2\ ,
\end{multline}}
\[
-(1-y)\log\left(1-\mathrm{pET}_{u,a}\right)-y \log \left(\mathrm{pET}_{u,a}\right)+\frac{\lambda}{2} \|\Theta\|_2^2\ ,
\]
$y \in \{0,1\}$ is the label (or indicator) for the positive event involving user $u$ and ad $a$, and $\lambda$ indicates the $L2$ regularization factor. The step sizes for OGD are determined using a variant of the AdaGrad (Adaptive Gradient) algorithm \cite{duchi2011adaptive}.

The \offset algorithm is based on incremental training, updating its model parameters with every batch of new training events, such as every 15 minutes for the click model. It includes an \textit{adaptive online hyper-parameter tuning mechanism} \cite{aharon2017adaptive} that utilizes the parallel \textit{Map-Reduce} architecture of the Gemini platform to adapt to changing market conditions, like trends and temporal effects. This mechanism tunes \offset's hyper-parameters, like the OGD initial step size and AdaGrad parameters. In addition, it also supports dynamic allocation of user feature LF sizes (implementing an internal automatic ``feature selection'' mechanism) and controls model size \cite{kaplan2021dynamic}.\naama{new, please revisit}\oren{done} See \cite{aharon2019soft} for other elements of \offset, such as the similarity weights for implementing ``soft'' recency and frequency regulations.

\Comment{
Other components of \offset, like the similarity weights for applying "soft" recency and frequency rules (how recent and how often a user has seen the same ad or campaign) \cite{aharon2019soft}, and a mechanism that chooses dynamically the length of the model's vector and the allocation of each independent and overlaps features  \cite{kaplan2021dynamic} are not included here for brevity.
}
\oren{new: I would add Rina's paper briefly here by saying that the hyper-parameter tuning is also powering the DVA and DVL. one sentence and say for more details see \cite{kaplan2021dynamic}}\naama{done}
\Comment{
\subsection{Weighted multi-value feature type}\label{sec:offset wmv feature}
\offset algorithm most generic feature type is the \textit{weighted multi-value} (WMV) feature type. To support this feature type, the model includes a $d$-dimension LF vector for each of the $m$ feature values encountered so far. Hence, the $d$-dimension vector of this WMV feature for user $u$ is\footnote{The definition is valid for ad entities as well, however, the feature values LF vectors and final vector in this case are of dimension $D$.} 
\[
v=\frac{1}{\sqrt{n}}\sum_{i=1}^n w_{\ell_i}\ v_{\ell_i}\ ,
\]
where $\{\ell_i\}$ and $\{w_{\ell_i}\}$ are the $n\le m$ feature values and accompanied weights associated with user $u$, and $\{v_{\ell_i}\}$ are the model LF vectors associated with values $\{\ell_i\}$. It is noted that the weights are given as part of the users' data and are not model parameters that are needed to be learned. Each time \offset encounters a new value for the feature, it assigns a random Gaussian vector to it with zero mean and covariance matrix $\eta\cdot I_d$, where $0\le \eta\ll 1$ and $I_d$ is the identity matrix.
It is noted that simpler feature types such as \textit{categorical features} (e.g., gender and age), or non-weighted multi-value features (e.g., all weights equal 1), may be seen as special cases of the WMV feature type.}

\Comment{
The \offset algorithm primarily uses the \textit{weighted multi-value feature} type. This means that for each of the $m$ feature values seen so far, the model has a $d$-dimensional LF vector associated with it. The resulting $d$-dimensional vector of the weighted multi-value feature for user $u$ is 
\[
v=\frac{1}{\sqrt{n}}\sum_{i=1}^n w_{\ell_i}\ v_{\ell_i}\ ,
\]
where ${\ell_i}$ are the $n$ feature values associated with user $u$, and ${w_{\ell_i}}$ are the weights accompanying these values. The model LF vectors associated with the values ${\ell_i}$ are ${v_{\ell_i}}$.
The same definition is also valid for ad entities, with the feature values' LF vectors and resulting vector having a dimension of $D$.

It's important to note that the weights are already given as part of the user data and do not need to be learned by the model. Whenever \offset encounters a new value for a feature, it assigns it a random Gaussian vector with zero mean and covariance matrix $\eta\cdot I_d$, where $0\le \eta\ll 1$ and $I_d$ is the identity matrix. This represents one of the ``cold-star'' strategies. Other, more advanced strategies are not covered in this work. It's also worth noting that simpler feature types, such as categorical features (e.g. age and gender) or non-weighted multi-value features (e.g. user category feature with all weights equal to 1), are just special cases of the weighted multi-value feature type.}
}
\section*{Acknowledgments}
\addcontentsline{toc}{section}{Acknowledgments}
The authors would like to thank Rotem Stram for providing insightful comments that significantly improved the quality of the manuscript.
\vspace{-0.1cm}
\balance
\Comment{
\bibliographystyle{plain}
\bibliography{references}
}


\end{document}